\documentclass[12pt,preprint]{aastex}

\def\eq#1{\begin{equation} #1 \end{equation}}

\def\comm#1           {{\tt (COMMENT: #1)}}

\received{2007 February 24}
\begin{document}

\title{             SDSS Standard Star Catalog for Stripe 82:
                  the Dawn of Industrial 1\% Optical Photometry       }
 
\author{
\v{Z}eljko Ivezi\'{c}\altaffilmark{\ref{Washington}},
J. Allyn Smith\altaffilmark{\ref{AustinPeay}},
Gajus Miknaitis\altaffilmark{\ref{FNAL}},
Huan Lin\altaffilmark{\ref{FNAL}},
Douglas Tucker\altaffilmark{\ref{FNAL}},
Robert H. Lupton\altaffilmark{\ref{Princeton}},
James E. Gunn\altaffilmark{\ref{Princeton}},
Gillian R. Knapp\altaffilmark{\ref{Princeton}}, 
Michael A. Strauss\altaffilmark{\ref{Princeton}},
Branimir Sesar\altaffilmark{\ref{Washington}},
Mamoru Doi\altaffilmark{\ref{UT}},
Masayuki Tanaka\altaffilmark{\ref{UT2}},
Masataka Fukugita\altaffilmark{\ref{UT3}},
Jon Holtzman\altaffilmark{\ref{NMSU}},
Steve Kent\altaffilmark{\ref{FNAL}},
Brian Yanny\altaffilmark{\ref{FNAL}},
David Schlegel\altaffilmark{\ref{LBNL}}, 
Douglas Finkbeiner\altaffilmark{\ref{Harvard}},
Nikhil Padmanabhan\altaffilmark{\ref{LBNL}},
Constance M. Rockosi\altaffilmark{\ref{UCSC}}, 
Mario Juri\'{c}\altaffilmark{\ref{IAS}},
Nicholas Bond\altaffilmark{\ref{Princeton}},
Brian Lee\altaffilmark{\ref{LBNL}},
Chris Stoughton\altaffilmark{\ref{FNAL}}, 
Sebastian Jester\altaffilmark{\ref{Southampton}}, 
Hugh Harris\altaffilmark{\ref{USNOFlagstaff}}, 
Paul Harding\altaffilmark{\ref{CWRU}}, 
Heather Morrison\altaffilmark{\ref{CWRU}}, 
Jon Brinkmann\altaffilmark{\ref{APO}}, 
Donald P. Schneider\altaffilmark{\ref{PennState}},
Donald York\altaffilmark{\ref{Chicago}}
}

\newcounter{address}
\setcounter{address}{1}
\altaffiltext{\theaddress}{University of Washington, Dept. of Astronomy, Box 351580, Seattle, WA 98195
\label{Washington}}
\addtocounter{address}{1}
\altaffiltext{\theaddress}{Dept. of Physics \& Astronomy, Austin Peay State University, Clarksville, TN 37044
\label{AustinPeay}}
\addtocounter{address}{1}
\altaffiltext{\theaddress}{Fermi National Accelerator Laboratory, P.O. Box 500, Batavia, IL 60510
\label{FNAL}}
\addtocounter{address}{1}
\altaffiltext{\theaddress}{Princeton University Observatory, Princeton, NJ 08544
\label{Princeton}}
\addtocounter{address}{1}
\altaffiltext{\theaddress}{Institute of Astronomy, University of Tokyo, 2-21-1 Osawa,
Mitaka, Tokyo 181-0015, Japan
\label{UT}}
\addtocounter{address}{1}
\altaffiltext{\theaddress}{Dept. of Astronomy, Graduate School of Science, University of Tokyo,
Hongo 7-3-1, Bunkyo-ku, Tokyo, 113-0033, Japan 
\label{UT2}}
\addtocounter{address}{1}
\altaffiltext{\theaddress}{
Institute for Cosmic Ray Research, University of Tokyo, Kashiwa, Chiba, Japan
\label{UT3}}
\addtocounter{address}{1}
\altaffiltext{\theaddress}{New Mexico State University, Box 30001, 1320 Frenger St., Las Cruces, NM 88003
\label{NMSU}}
\addtocounter{address}{1}
\altaffiltext{\theaddress}{Lawrence Berkeley National Laboratory, One Cyclotron Road, MS 50R5032, Berkeley, CA, 94720 
\label{LBNL}}
\addtocounter{address}{1}
\altaffiltext{\theaddress}{Harvard-Smithsonian Center for Astrophysics, 60 Garden Street,
                          Cambridge, MA 02138
\label{Harvard}}
\addtocounter{address}{1}
\altaffiltext{\theaddress}{University of California--Santa Cruz, 1156 High St., Santa Cruz, CA 95060
\label{UCSC}}
\addtocounter{address}{1}
\altaffiltext{\theaddress}{Institute for Advanced Study, 1 Einstein Drive, Princeton, NJ 08540
\label{IAS}}
\addtocounter{address}{1}
\altaffiltext{\theaddress}{School of Physics and Astronomy, University of
               Southampton, Highfield, Southampton, SO17 1BJ, UK
\label{Southampton}}
\addtocounter{address}{1}
\altaffiltext{\theaddress}{U.S. Naval Observatory, Flagstaff Station, P.O. Box 1149, Flagstaff, AZ 86002
\label{USNOFlagstaff}}
\addtocounter{address}{1}
\altaffiltext{\theaddress}{Department of Astronomy, Case Western Reserve University, Cleveland, Ohio 44106 
\label{CWRU}}
\addtocounter{address}{1}
\altaffiltext{\theaddress}{Apache Point Observatory, 2001 Apache Point Road, P.O. Box 59, Sunspot, NM 88349-0059
\label{APO}}
\addtocounter{address}{1} \altaffiltext{\theaddress}{Department of Astronomy
and Astrophysics, The Pennsylvania State University, University Park, PA 16802
\label{PennState}}
\addtocounter{address}{1}
\altaffiltext{\theaddress}{University of Chicago, Astronomy \& Astrophysics Center, 5640 S. Ellis Ave., Chicago, IL 60637
\label{Chicago}}

\begin{abstract}
We describe a standard star catalog constructed using multiple SDSS photometric 
observations (at least four per band, with a median of ten) in the $ugriz$ 
system. The catalog includes 1.01 million non-variable unresolved objects from the 
equatorial stripe 82 ($|\delta_{J2000}|<$ 1.266$^\circ$) in the RA range 20h 34m to 4h 00m, 
and with the corresponding $r$ band (approximately Johnson V band) magnitudes 
in the range 14--22. The distributions of measurements 
for individual sources demonstrate that the photometric pipeline correctly estimates 
random photometric errors, which are below 0.01 mag for stars brighter than (19.5, 
20.5, 20.5, 20, 18.5) in $ugriz$, respectively (about twice as good as for 
individual SDSS runs). Several independent tests of the internal consistency
suggest that the spatial variation 
of photometric zeropoints is not larger than $\sim$0.01 mag (rms). In addition to being
the largest available dataset with optical photometry internally consistent at the
$\sim$1\% level, this catalog provides practical definition of the SDSS 
photometric system. Using this catalog, we show that photometric zeropoints
for SDSS observing runs can be calibrated within nominal uncertainty of 2\% even
for data obtained through 1 mag thick clouds, and demonstrate the existence of 
He and H white dwarf sequences using photometric data alone. Based on the 
properties of this catalog, we conclude that upcoming large-scale optical
surveys such as the Large Synoptic Survey Telescope will be capable of
delivering robust 1\% photometry for billions of sources.  
\end{abstract}


\section{                          Introduction                     }

Astronomical optical photometric data are usually calibrated using sets of
standard stars 
whose brightness is known from previous work. The most notable modern optical 
standard star catalogs are Landolt standards (Landolt 1992) and Stetson standards
(Stetson 2000, 2005). Both are reported on the Johnson-Kron-Cousins system (Landolt
1983 and references therein). 
The Landolt catalog provides magnitudes accurate to 1-2\% in the $UBVRI$ bands for $\sim$500
stars in the $V$ magnitude range 11.5--16. Stetson has extended Landolt's work to 
fainter magnitudes, and provided the community with $\sim$1-2\% accurate magnitudes 
in the $BVRI$ bands for $\sim$15,000 stars in the magnitude range $V\la20$. Most 
stars from both sets are distributed along the Celestial Equator, which facilitates 
their use from both hemispheres.

The data obtained by the Sloan Digital Sky Survey (SDSS; York et al. 2000) can be 
used to extend the work by Landolt and Stetson to even fainter levels, and to increase 
the number of standard stars to over a million. In addition, SDSS has designed its own 
photometric system ($ugriz$, Fukugita et al. 1996) which is now in use at many
observatories worldwide. This widespread use of the $ugriz$ photometric 
system motivates the construction of a large standard star catalog with $\sim$1\% 
accuracy. As a part of its imaging survey, SDSS has obtained many scans in the 
so-called Stripe 82 region, which is defined by $|\delta_{J2000}|<$ 1.266$^\circ$ and 
RA approximately in the range 20h -- 4h. These repeated observations can be averaged 
to produce more accurate photometry than the nominal 2\% single-scan accuracy 
(Ivezi\'{c} et al. 2004a).  

The catalog and methods presented here have some similarity with an effort
by Padmanabhan et al. (2007), who developed a new calibration algorithm 
that simultaneously solves for the calibration parameters and relative stellar 
fluxes using overlapping SDSS observations (the so-called {\it \"{u}bercalibration} 
method). The algorithm decouples the problem  of ``relative'' calibrations
(i.e. producing an internally consistent system), from that of ``absolute'' 
calibrations (i.e. typing the internal system to a physical flux scale):
the absolute calibration is reduced to determining a few numbers for the 
entire survey. Here we also decouple ``relative'' and ``absolute''
calibrations and use overlapping observations. The main difference between 
their work and this paper is that they are concerned about calibrating
the entire SDSS survey ($\sim$8500 deg$^{2}$; most of the surveyed area has 
at most two overlapping observations), while we concentrate here on a much 
smaller area ($\sim$300 deg$^{2}$) with an average of ten overlapping
observations. We also determine flatfield corrections (relative to the
``standard'' survey reductions) using different methods: Padmanabhan et al.
minimize errors in relative photometry of multiply observed stars, while
we require that the stellar locus remains fixed in multi-dimensional 
color space. An advantage of the catalog presented here is better 
averaging of various photometric errors thanks to a larger number of observations,
which comes at the expense of a much smaller cataloged area. It is encouraging 
that the results of these two complementary approaches agree in the regions 
of sky common to both catalogs at the claimed level of accuracy ($\sim$1\%).

Additional motivation for the
analysis of repeated scans and their impact on photometric accuracy comes
from upcoming large-scale optical surveys such as the Dark Energy Survey
(Flaugher et al. 2007), Pan-STARRS (Kaiser 2002) and the Large Synoptic Survey
Telescope (Tyson 2002, LSST hereafter). 
For example, the LSST science requirements document\footnote{Available from
http://www.lsst.org/Science/lsst\_baseline.shtml} 
calls for a photometric system 
that is internally consistent across the sky at the 1\% level. The SDSS Stripe
82 repeated scans can be used to gauge the plausibility of delivering such a 
system. 

We describe the construction and testing of a standard star catalog in
\S 2, and illustrate its several use cases in \S 3. We discuss our results 
in \S 4.

\section{ The Construction of SDSS Stripe 82 Standard Star Catalog }

\subsection{             Overview of SDSS imaging data                }
\label{overview}
SDSS is using a dedicated 2.5m telescope (Gunn et al. 2006) to provide homogeneous 
and deep ($r < 22.5$) photometry in five bandpasses 
(Fukugita et al.~1996; Gunn et al.~1998, 2006; Smith 
et al.~2002; Hogg et al. 2002) repeatable to 0.02 mag (root-mean-square scatter,
hereafter rms, for sources not limited by photon statistics, Ivezi\'{c} et al.~2003) 
and with a zeropoint uncertainty of $\sim$0.02-0.03 (Ivezi\'{c} et al.~2004a). 
The survey sky coverage of close to $\sim$10,000 deg$^2$ 
in the Northern Galactic Cap, and $\sim$300 deg$^2$ in the Southern Galactic 
Hemisphere, will result in photometric measurements for well over 100 million stars and 
a similar number of galaxies\footnote{The recent Data Release 5
(Adelman-McCarthy et al. 2007, in prep) lists photometric
data for 215 million unique objects observed in 8000 deg$^2$ of sky;
see http://www.sdss.org/dr5/.}. Astrometric positions are accurate to better than 
0.1 arcsec per coordinate (rms) for sources with $r<20.5^m$ (Pier et al.~2003), and 
the morphological information from the images allows reliable star-galaxy separation 
to $r \sim$ 21.5$^m$ (Lupton et al.~2002, Scranton et al. 2002). 

Data from the imaging camera (thirty photometric, twelve astrometric, and two focus CCDs, 
Gunn et al. 1998) are collected in drift scan mode. The images that correspond 
to the same sky location in each of the five photometric bandpasses (these five images 
are collected over $\sim$5 minutes, with 54 sec for each exposure) are grouped 
together for simultaneous processing as a field. A field is defined as a 36
seconds (1361 pixels, or 9 arcmin, see Stoughton et al. 2002) long and 2048
pixels wide (13 arcmin) stretch of drift-scanning data from a single column of
CCDs (sometimes called
a scanline, for more details please see Stoughton et al. 2002; Abazajian et
al. 2003, 2004, 2005; Adelman-McCarthy et al. 2006). Each of the six scanlines 
(called together a strip) is 13 arcmin wide. The twelve interleaved scanlines
(or two strips) are called a stripe ($\sim$2.5$^\circ$ wide).

\subsection{ The photometric calibration of SDSS imaging data  }

\label{photomCalib}

SDSS 2.5m imaging data are photometrically calibrated using a network of calibration stars 
obtained in 1520 41.5$\times$41.5 arcmin$^2$ transfer fields, called secondary patches.
These patches are positioned throughout the survey area and are calibrated using a primary 
standard star network of 158 stars distributed around the Northern sky (so-called USNO 
standards, Smith et al.~2002). 
The primary standard star network is tied to an absolute flux system by the single F0 subdwarf 
star BD+17$^\circ$4708, whose absolute fluxes in SDSS filters are taken from Fukugita et al. (1996).
The secondary patches are grouped into sets of four, and are observed by the
Photometric Telescope (hereafter PT; Tucker et al. 2006) in parallel with observations of the primary standards.  
A set of four patches spans all 12 scanlines of a survey stripe along the width of the stripe, 
and the sets are spaced along the length of a stripe at roughly 15 degree intervals,
which corresponds to an hour of scanning at the sidereal rate. 

SDSS 2.5m magnitudes are reported on the "natural system" of the 2.5m telescope defined by 
the photon-weighted effective wavelengths of each combination of SDSS filter, CCD response, 
telescope transmission, and atmospheric transmission at a reference airmass of 1.3 as 
measured at APO\footnote{Transmission curves for the SDSS 2.5m photometric system are
available at http://www.sdss.org/dr5/instruments/imager}. The magnitudes are referred
to as the $ugriz$ system (which differs from the ``primed'' system, $u'g'r'i'z'$, that 
is defined by the USNO standards\footnote{For subtle effects that led to this distinction,
please see Stoughton et al. (2002), Smith et al. (2002), and 
http://www.sdss.org/dr5/algorithms/fluxcal.html.}). The reported 
magnitudes\footnote{SDSS uses a modified magnitude system (Lupton, Szalay \& Gunn 1999),
which is virtually identical to the standard astronomical Pogson magnitude system at 
high signal-to-noise ratios relevant here.} are corrected for the atmospheric extinction 
(using simultaneous observations of standard stars by the PT) and thus 
correspond to measurements at the top of the atmosphere\footnote{The same atmospheric 
extinction correction is applied irrespective of the source color; the systematic errors
this introduces are probably less than 1\% for all but objects of the most extreme
colors.} 
(except for the fact that the 
atmosphere has an impact on the wavelength dependence of the photometric system response). 
The magnitudes are reported on the AB system (Oke \& Gunn  1983) defined such 
that an object with a specific flux of $F_\nu$=3631 Jy  has $m=0$ (i.e. an object with 
$F_{\nu}$=const. has an AB magnitude equal to the Johnson $V$ magnitude at all wavelengths).
In summary, given a specific flux of an object {\it at the top} of the atmosphere, $F_\nu(\lambda)$,
the reported SDSS 2.5m magnitude in a given band, $b$=($u, g, r, i, z$),
corresponds to (modulo random and systematic errors, which will be discussed later)
\begin{equation}
\label{ABmag}
       m = -2.5\log_{10}\left({F_b \over 3631 \, {\rm Jy}}\right),
\end{equation}
where 
\begin{equation}
      F_b = \int{F_\nu(\lambda) \phi_b(\lambda) d\lambda}.
\end{equation}
Here, $\phi_b(\lambda)$ is the normalized system response for the given band,
\begin{equation}
\label{PhiDef}
      \phi_b(\lambda) = {\lambda^{-1} S_b(\lambda) \over \int{\lambda^{-1} S_b(\lambda) d\lambda}},
\end{equation}
with the overall atmosphere + system throughput, $S_b(\lambda)$, available from the website 
given above ($\phi_b(\lambda)$ for the SDSS system are shown in
Figure~\ref{compareBandpasses}, see also \S~\ref{sixSystems}). We reiterate that the 
{\it normalization} of reported magnitudes corresponds to a source at the top of the atmosphere, 
while the throughput $\phi_b(\lambda)$ includes the transmission of a standard atmosphere at a fiducial
airmass of 1.3. Note also that it is only the {\it shape} of $S_b(\lambda)$,
and not its overall normalization, that needs to be known to compute 
expected SDSS magnitudes of a source with given $F_\nu(\lambda)$. That is,
the SDSS photometric system is fully {\it defined} by the five {\it dimensionless}
functions $\phi_b(\lambda)$ (by definition, $\int\phi_b d\lambda=1$,
see eq.~\ref{PhiDef}). In reality, for each $ugriz$ band there are six devices
in the SDSS camera (Gunn et al. 1998) whose  $\phi_b$ are slightly different
(see \S~\ref{sixSystems}).

The quality of SDSS photometry stands out among available large-area optical sky surveys 
(Ivezi\'{c} et al.~2003, 2004a; Sesar et al.~2006). Nevertheless, the achieved accuracy is 
occasionally worse than the nominal 0.02-0.03 mag (root-mean-square scatter for sources not limited 
by photon statistics). Typical causes of substandard photometry include an incorrectly 
modeled point spread function (PSF; usually due to fast variations of atmospheric seeing, or lack of a sufficient 
number of the isolated bright stars needed for modeling the PSF), unrecognized changes in atmospheric 
transparency, errors in photometric zeropoint calibration, effects of crowded fields 
at low Galactic latitudes, an undersampled PSF in excellent seeing conditions ($\la 0.8$ 
arcsec; the pixel size is 0.4 arcsec), incorrect flatfield or bias vectors, scattered 
light erroneously included in flatfield, etc. Such effects can conspire to
increase the photometric errors to 
levels as high as 0.05 mag (with a frequency, at that error level, of roughly one field 
per thousand). However, when multiple scans of the same sky region are available, many 
of these errors can be minimized by properly averaging photometric measurements.

\subsection{           The Choice of Cataloged Magnitudes             }
\label{mags} 

The SDSS photometric pipeline ({\it photo}, Lupton et al. 2002) measures several
types of magnitudes, including aperture, PSF, and model magnitudes. Here we
briefly describe each type of magnitude (for more details see Stoughton et
al. 2002, and the SDSS website www.sdss.org) and justify the choice of PSF 
magnitudes for catalog construction.

\subsubsection{ Aperture magnitudes }

Aperture magnitudes computed by {\it photo} are based on the flux contained 
within the aperture with a radius of 7.43 arcsec. While an aperture magnitude
is the most robust flux estimate at the bright end (because it is essentially
seeing independent), these magnitudes do not have good noise properties at 
the faint end where sky noise dominates (e.g. for a given maximum photometric 
error, PSF magnitudes reach 1--1.5 mag fainter than do aperture magnitudes). 
In order to improve the depth of the standard star catalog, we opt not to use
aperture magnitudes, except for quality tests at the bright end.

\subsubsection{ Point spread function magnitudes }

The point spread function (PSF) flux is computed using the PSF as a weighting 
function. While this flux is optimal for faint point sources (in particular,
it is vastly superior to aperture photometry at the faint end), it is also
sensitive to inaccurate PSF modeling as a function of position and time.
Even in the absence of atmospheric variations, the SDSS telescope and camera
optics deliver images whose FWHMs vary by up to 15\% from one side 
of a CCD to the other; the worst effects are seen in the chips farthest from the 
optical axis. Moreover, since the atmospheric seeing varies with time, the delivered 
image quality is a complex two-dimensional function even on the scale of a single frame. 
Without accounting for this spatial variation, the PSF photometry would have
errors up to 0.10-0.15 mag. The description of the PSF is
also critical for star-galaxy separation and for unbiased measures of the
shapes of nonstellar objects.

The SDSS imaging PSF is modeled heuristically in each band and each camera
column using a Karhunen-Lo\'{e}ve (KL) transform (Lupton et al. 2002). Using
stars brighter than roughly 20$^{\rm th}$ magnitude, the PSF from a series of
five frames is expanded into eigenimages and the first three terms are
retained. The variation of these coefficients is then fit up to a second order 
polynomial in each chip coordinate. The failure of this KL expansion,
typically due to insufficient number of stars, or exceedingly complex 
PSF, results in occasional problems with PSF photometry. The main failure
mode is inaccurate determination of aperture corrections which statistically
tie PSF magnitudes to aperture magnitudes using bright stars.

\subsubsection{   Model magnitudes }

Just as the PSF magnitudes are optimal measures of the fluxes of stars, the optimal 
measure of the flux of a galaxy uses a matched galaxy model. With this in mind, 
the photometric pipeline fits two models to the two-dimensional image of each object 
in each band: a pure deVaucouleurs profile and a pure exponential profile\footnote{For
more details see http://www.sdss.org/dr5/algorithms/photometry.html}. 
Because the models are convolved with a double-Gaussian fit to the PSF, the seeing
effects are accounted for. Aperture corrections are applied to make these model 
magnitudes equal the PSF magnitudes in the case of an unresolved object.

\subsubsection{ The choice of magnitudes for the standard star catalog  } 
\label{magTypes}
A comparison between aperture, PSF and model magnitudes for unresolved sources
is done automatically for every SDSS observing run ({\it runQA} pipeline, Ivezi\'{c}
et al. 2004a). An analysis of over 200 runs indicate that model magnitudes
are more robust than PSF magnitudes: PSF magnitudes show systematic offsets
of 0.05 mag from aperture magnitudes three times more often than do model 
magnitudes (roughly once per thousand fields), as model fits have more degrees 
of freedom. On the other hand, an
analysis of repeated scans indicates that estimates of photometric errors
by the photometric pipeline are more accurate for PSF magnitudes (agreeing at
the 10\% level with the measured values, see Ivezi\'{c} et al. 2003; 
Scranton et al. 2005) than for model magnitudes
(which are smaller than the measured values by typically 30-50\%). Because
the rejection of likely variable sources, which relies on accurate photometric
error estimates, is an important step in the
construction of the standard star catalog (see below), we choose to 
use PSF magnitudes to construct the catalog.

\subsection{                Catalog Construction }

Using 58 SDSS-I runs from stripe 82 (approximately 20h $< \alpha_{J2000} <$ 04h
and $|\delta_{J2000}|<$ 1.266$^\circ$, but not all runs extend over the entire right
ascension range) 
obtained in mostly photometric conditions (as 
indicated by the PT calibration residuals, infrared cloud camera\footnote{For more
details about the camera see http://hoggpt.apo.nmsu.edu/irsc/irsc\_doc 
and Hogg et al. (2002).}, 
and tests performed by the {\it runQA} pipeline),
candidate standard stars from each run are selected by requiring  
\begin{enumerate}
\item 
that objects are classified as STAR (based on the difference between model
and PSF magnitudes); this morphological classification really means unresolved 
(point) sources (e.g. quasars are also included),
\item
that they have quoted photometric errors in the PSF magnitude (as computed by
the photometric pipeline)
smaller than 0.05 mag in at least one band, and 
\item
that the processing flags BRIGHT, SATUR, BLENDED, or EDGE are not set
in any band\footnote{For
more details about processing flags see 
http://www.sdss.org/dr5/products/catalogs/flags.html
and Stoughton et al. (2002).}. 
\end{enumerate}
These criteria select unsaturated point sources with sufficiently
high signal-to-noise ratio per single observation to approach final 
photometric errors of 0.02 mag or smaller.

After matching all detections of a single source across runs (using a 1 arcsec
matching radius), various photometric statistics such as unweighted mean, median, 
their standard errors, 
root-mean-square scatter, number of observations, and $\chi^2$ per degree 
of freedom, are computed for magnitudes in each band. We use
errors reported by the photometric pipeline to compute $\chi^2$ and note
that systematic errors common to all runs do not contribute to its value.
This initial 
catalog of multi-epoch observations includes 1.4 million point sources with at least 
4 observations in each of the $g$, $r$ and $i$ bands. The median number of 
observations per source and band is 10, and the total number of photometric 
measurements is $\sim$57 million. 

\begin{figure}
\epsscale{.8}
\plotone{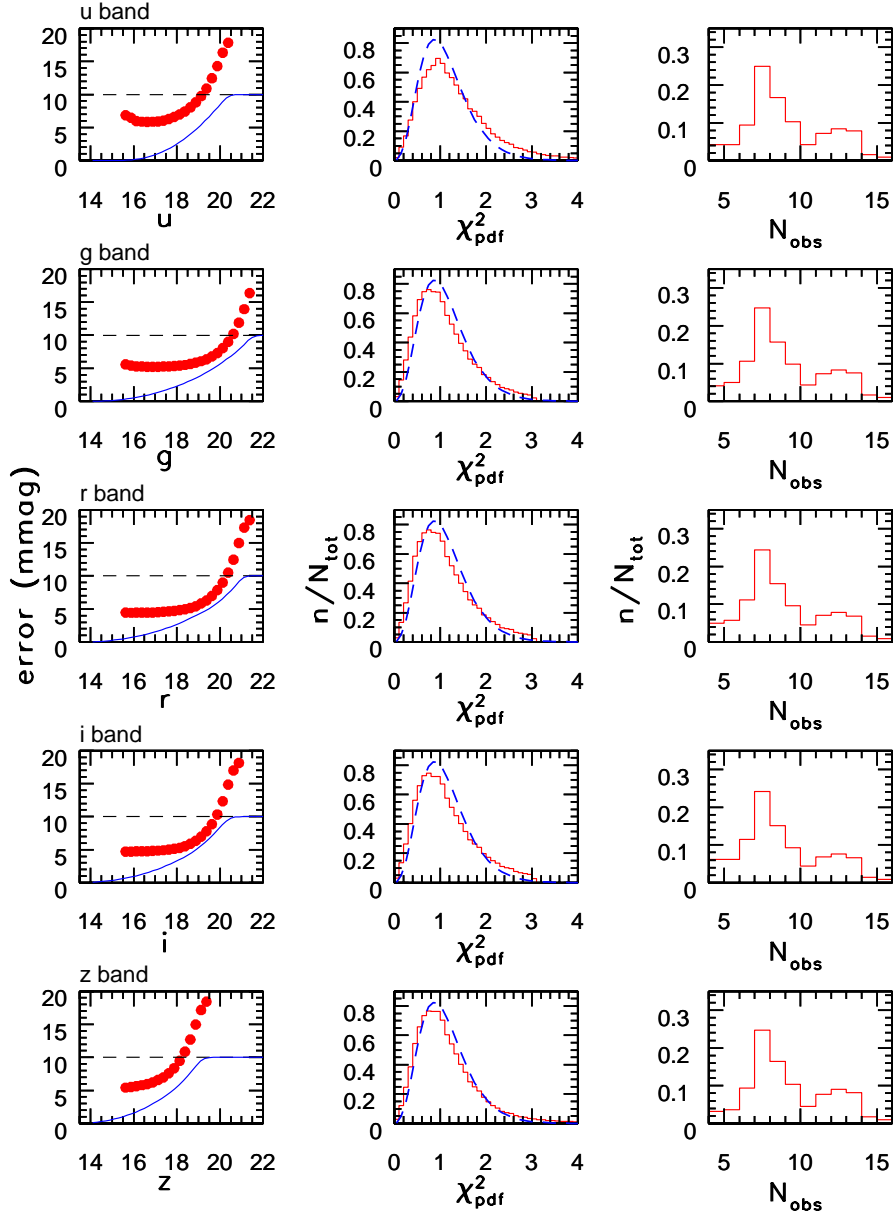}
\caption{\small The median magnitude error as a function of magnitude
(left column, symbols), the $\chi^2$ per degree of freedom distribution
(solid line, middle column), and the number of observations in 
each band  (right column) for candidate standard stars from 
SDSS stripe 82 (selected by $\chi^2<3$ in the $gri$ bands). 
The solid lines in the left column show cumulative magnitude
distribution normalized to 10 at the faint end. 
The dashed lines in the middle column show the $\chi^2$ per degree 
of freedom distribution for a Gaussian error distribution and 
nine degrees of freedom. Its similarity with the measured distributions 
suggests that the magnitude errors computed by the photometric 
pipeline are reliable (they may be slightly underestimated, by 
about 10\%, in the $u$ band).
\label{multiData}}
\end{figure}

\begin{figure}
\plotone{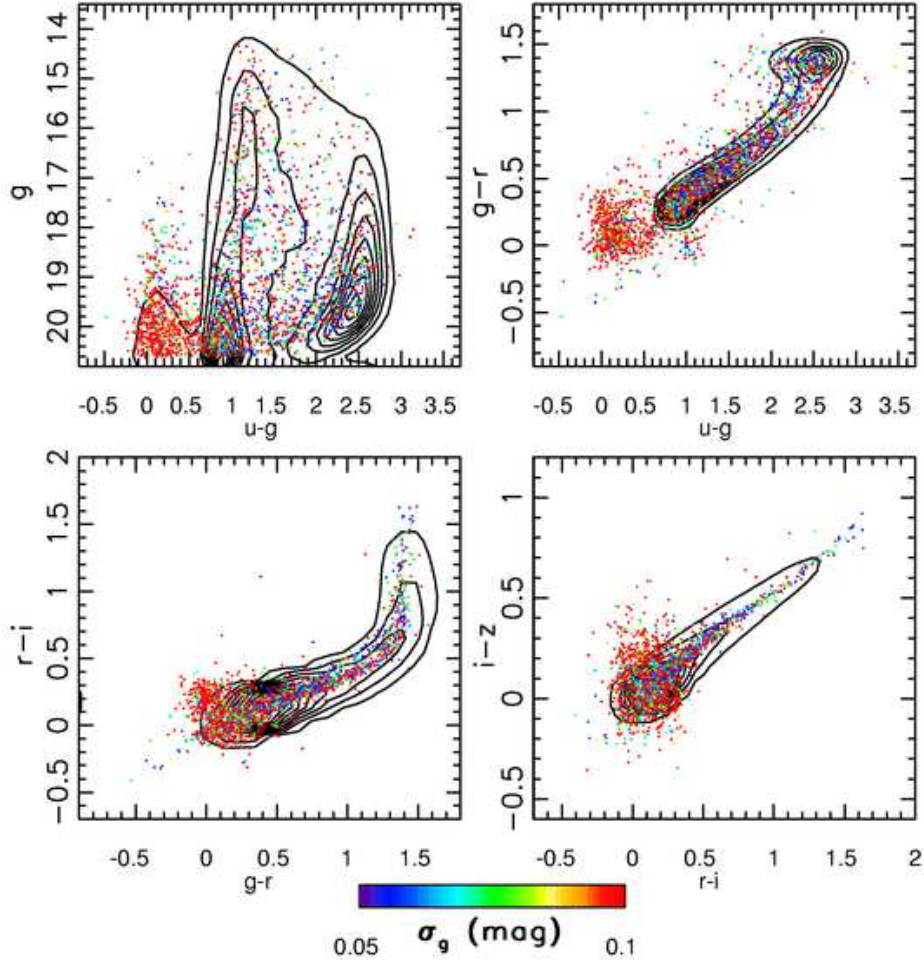}
\caption{\small A comparison of the color-magnitude and color-color distributions
for variable and non-variable unresolved sources from SDSS stripe 82.
The distributions of sources with a root-mean-square (rms) scatter 
of the $g$ band magnitude below 0.05 mag are shown by linearly-spaced 
contours (non-variable sources). Sources with $320^\circ <$ RA $< 330^\circ$, 
with rms scatter in the $g$ band larger than 0.05 mag, and $\chi^2$ greater
than three, are shown by dots (variable sources). The dots are color-coded 
according to the observed rms scatter in the $g$ band (0.05--0.10 mag, see 
the legend, red indicates variability larger than 0.1 mag). Note how 
low-redshift quasars ($u-g<0.6$) and RR Lyrae ($u-g\sim1.1$, $g-r\sim0$; 
Ivezi\'{c} et al. 2005) clearly stand out as variable sources (red points).  
\label{CMD1}}
\end{figure}

The distributions of the median magnitudes, their standard errors, $\chi^2$ 
and the number of observations for a subset of these sources are shown in 
Figure~\ref{multiData}. The random errors in the 
median magnitude (computed as 0.928*IQR/$\sqrt{N-1}$, where IQR is the 
25\%--75\% interquartile range of the individual measurement distribution and
$N$ is the number of measurements; note that the error of the median for 
a Gaussian distribution is 25\% larger than the error of the mean, Lupton
1993) are below 0.01 
mag at the bright end. These errors are reliably computed by the photometric pipeline, 
as indicated by the $\chi^2$ distributions. The distributions of these sources
in color-magnitude and color-color diagrams, constructed using median
magnitudes, are shown in Figure~\ref{CMD1}. For a detailed interpretation 
of these diagrams, please see Lenz et al. (1998), Fan (1999), Finlator et 
al. (2000), Helmi et al. (2003), and Ivezi\'{c} et al. 2006. As evident,
the sample is dominated by stars.

\subsubsection{       Selection of Candidate Standard Stars         }

\begin{figure}
\epsscale{.9}
\plotone{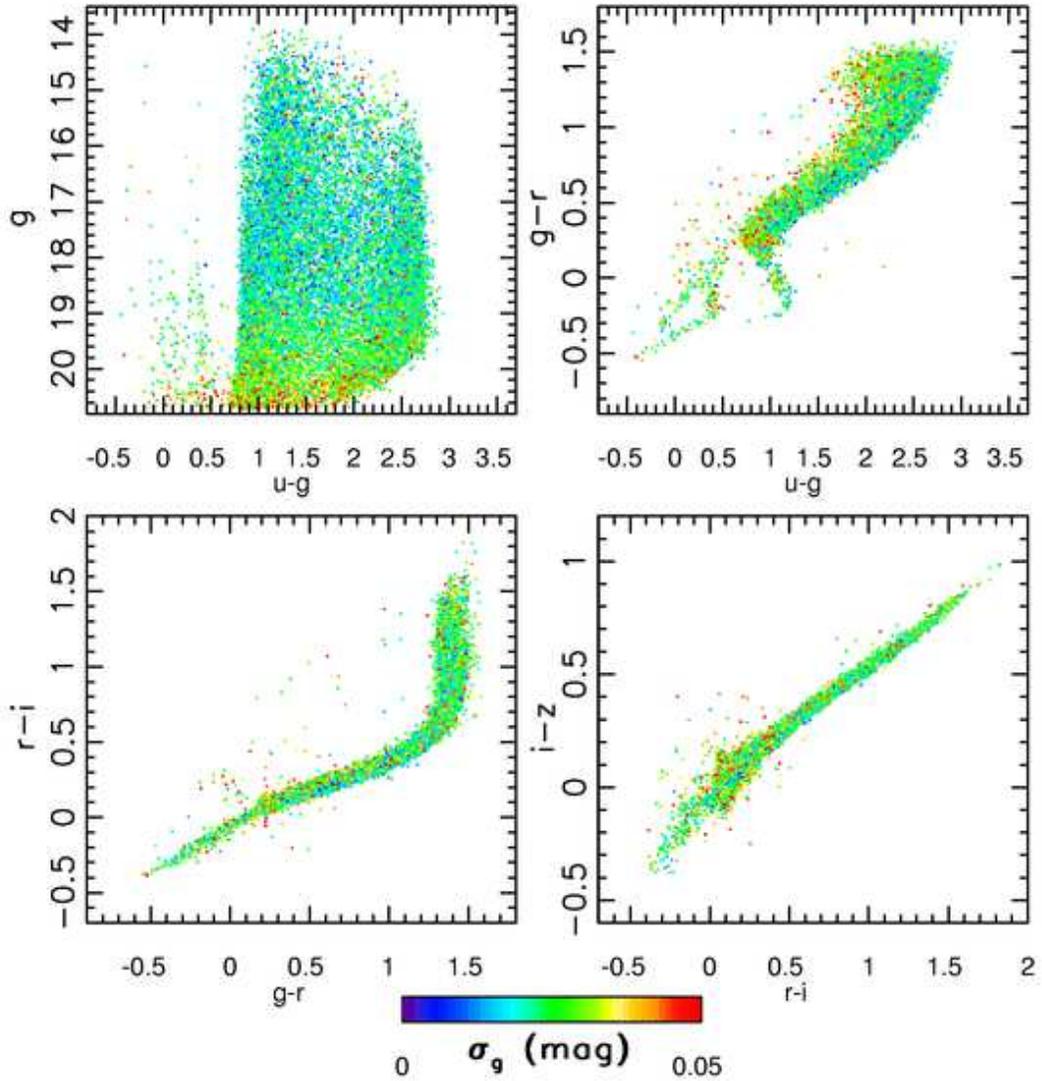}
\caption{The color-magnitude and color-color distributions
of candidate standard stars brighter than $g=20$. The dots are 
color-coded according to the observed rms scatter in the 
$g$ band (0--0.05 mag, see the legend). Note the absence of 
quasars and RR Lyrae visible in Figure~\ref{CMD1}. 
\label{CMD2}}
\end{figure}

Adopted candidate standard stars must have at least 4 observations and, 
to avoid variable sources, $\chi^2$ less than 3 in the $gri$ bands
(the same requirements are later applied in the $u$ and $z$ bands when
using the catalog for calibration, as we discuss further below). We also
limit the right ascension to the range from 20h 34m to 4h 00m, which provides a simple
areal definition (together with $|\delta_{J2000}|<$1.266$^\circ$) of a
282 deg$^2$ large rectangular region, while excluding only a 
negligible fraction of stars. With the final condition that the standard 
error for the mean magnitude in the $r$ band is smaller than 0.05 mag, these 
requirements result in a catalog with
slightly over a million sources (1,006,849) with 42 million photometric 
measurements. Of those, 563,908 have a random error for the median magnitude 
in the $r$ band smaller than 0.01 mag, and for 405,954 stars this is true
in all three of the $gri$ bands. Subsets of 92,905 and 290,299 stars satisfy 
these requirements in the $ugri$ and $griz$ bands, and 91,853 stars satisfy
this in all five bands. 
The distributions of candidate standard stars that satisfy the above 
selection criteria in all five bands in color-magnitude and color-color 
diagrams are shown in Figure~\ref{CMD2}. 

\begin{figure}
\plotone{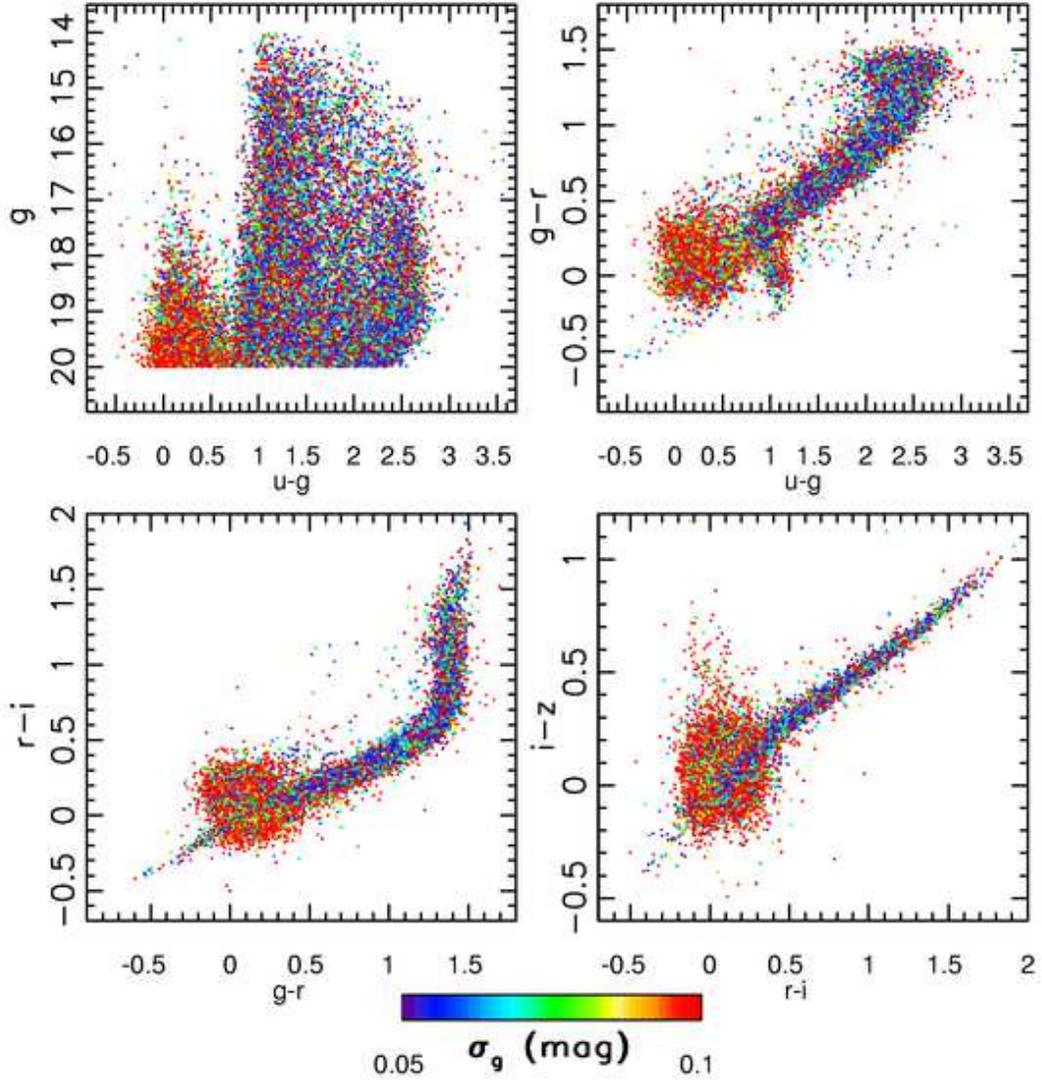}
\caption{Analogous to Figure~\ref{CMD2}, except that only 
variable sources are shown, and with a different color coding 
(0.05--0.10 mag range of the $g$ band rms, instead of 0--0.05
mag). 
\label{CMD3}}
\end{figure}

For comparison, the distribution of sources that were rejected as variable 
($\chi^2$ greater than 3 in at least one of the $gri$ bands)
in color-magnitude and color-color diagrams is shown in Figure~\ref{CMD3}. 
As evident from a comparison with Figure~\ref{CMD2}, the distribution of 
variable sources in color-color diagrams is markedly different from that 
of non-variable sources. It is especially striking how low-redshift ($z<2.2$)
quasars are easily detected by their variability (for more details see
Ivezi\'{c} et al. 2004b). However,
it is fairly certain that not all variable sources are recognized as such because
of the limited number of repeated observations ($\sim$10). For example, 
an eclipsing binary with much shorter eclipse duration than the orbital
period could easily escape detection. Analysis of the variable subsample
is presented in a companion paper (Sesar et al. 2007, in prep.). 

\begin{figure}
\plotone{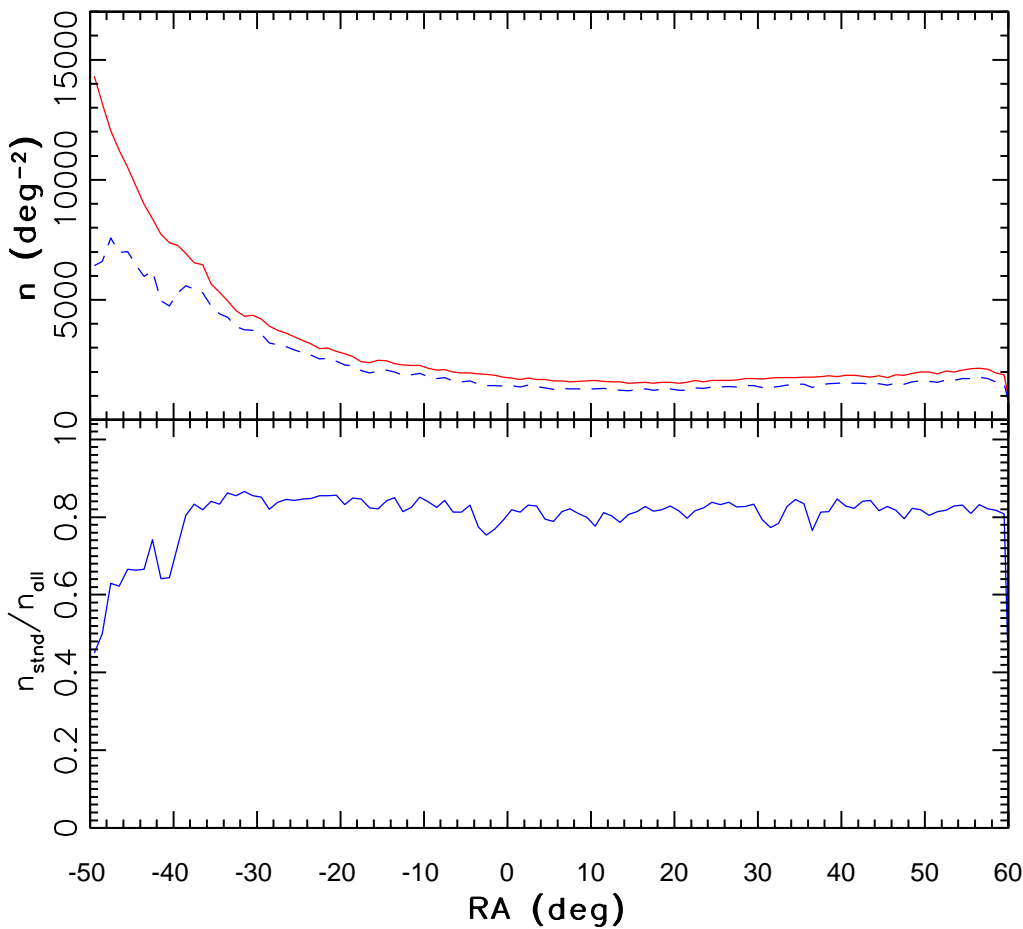}
\vspace{-1.0in}
\caption{The top panel shows the sky density of all the point sources 
with at least 4 observations in each of the $g$, $r$ and $i$ bands (solid),
and of selected candidate standard (non-variable) stars (dashed). The
bottom panel shows the ratio of the two curves shown in the top panel.
For reference, Galactic coordinates, ($l$,$b$), are (46,$-$24), (96,$-$60)
and (190,$-$37) for $\alpha_{J2000}$=$-$50$^\circ$, 0$^\circ$ and 60$^\circ$ 
(at $\delta_{J2000}$=0$^\circ$).
\label{RAcounts}}
\end{figure}

\begin{figure}
\plotone{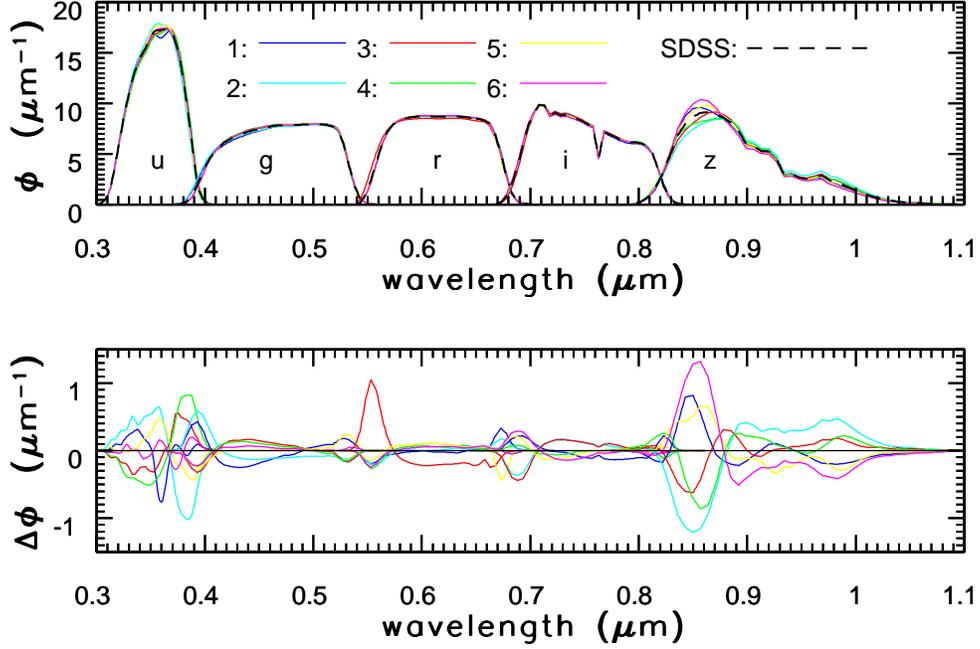}
\vspace{-2cm}
\vspace{-1.0in}
\caption{The top panel compares the five transmission curves, $\phi$ 
(see eq.\ref{PhiDef} for definition), that define the SDSS photometric 
system (dashed lines), to measured transmission curves sorted by camera 
column and shown by solid lines, as marked in the panel (the area under each 
curve is unity by definition). The bottom panel shows the differences between
the measured transmission curves and the curves that define the system.
These differences induce color terms that result in systematic photometric
errors as a function of source color. The largest color terms are present for
the $z$ band in camera columns 2 and 6, with the errors well described
by $\Delta z_2 = z_2 - z_{\rm SDSS} = -0.019(r-i)$ and $\Delta z_6 =
+0.017(r-i)$ mag, respectively.
\label{compareBandpasses}}
\end{figure}

The sky density of all the sources and those selected as non-variable are
shown in Figure~\ref{RAcounts}. At high Galactic latitudes ($|b|\sim60^\circ$) 
the fraction of candidate standard stars sources is $\sim$80\%.

\subsection{              Systematic Photometric Errors              }

Photometric errors computed by the photometric pipeline provide a good
estimate of random errors in SDSS photometry, as demonstrated by the 
$\chi^2$ distributions shown in Figure~\ref{multiData}. However, the 
measurements are also subject to systematic errors such as spatial 
dependence of the internal zeropoints (calibration errors), and the 
overall deviations of the internal SDSS zeropoints from an AB magnitude 
scale. Formally, the true AB magnitude of an object (defined by 
eq.\ref{ABmag}) in a given band, $m_{true}$, can be expressed as 
\eq{
\label{photoSysErr}
       m_{true} = m_{cat} + \delta_m({\rm RA,Dec}) + \Delta_m,
}
where $m_{cat}$ is the cataloged magnitude, $\delta_m({\rm RA,Dec})$ describes
the spatial variation of the internal zeropoint error around $\Delta_m$ (thus the
average of $\delta_m$ over the cataloged area is 0 by construction), and $\Delta_m$ 
is the overall (spatially independent) deviation of the internal SDSS 
system from a perfect AB system (the five values $\Delta_m$  
are equal for all the cataloged objects). Here we ignore 
systematic effects, e.g., device non-linearity and bandpass variations 
between different camera columns, which depend on individual source 
properties such as brightness and colors (but see \S \ref{sixSystems} 
below).

The spatial variation of the internal zeropoint error can be separated 
into {\it color} errors, relative to a fiducial band, say $r$, and an 
overall ``gray'' error (e.g. unrecognized temporal changes in 
atmospheric transparency due to gray clouds)
\eq{
  \delta_m({\rm RA,Dec}) = \delta_r({\rm RA,Dec}) + \delta_{mr}({\rm RA,Dec}).
}
Below we discuss methods for estimating both the ``gray'' error 
$\delta_r({\rm RA,Dec})$ and the color errors $\delta_{mr}({\rm RA,Dec})$.

The deviation of the internal SDSS system from a perfect AB system,
$\Delta_m$, can also be expressed relative to the fiducial $r$ band
\eq{
       \Delta_m = \Delta_r + \Delta_{mr}.
}  
The motivation for this separation is twofold. First, $\Delta_{mr}$ can
be constrained by considering the colors (spectral energy distributions) 
of objects, independently from the overall flux scale (this can be done
using both external observations and models). Second, it is difficult
to find a science result that crucially depends on knowing the ``gray 
scale'' offset, $\Delta_r$, at the 1-2\% level. On the other hand, 
knowing the ``band-to-band'' offsets, $\Delta_{mr}$, with such an accuracy
is important for many applications (e.g., photometric redshifts of galaxies, 
type Ia supernovae cosmology, testing of stellar and galaxy models). 

Fitting SDSS spectra of hot white dwarfs to models, Eisenstein et al. (2006) 
determined the AB color corrections $\Delta_{mr}$ to be $-0.04$, $0.00$, $-0.015$ 
and $-0.015$ mag for $m=ugiz$, respectively, with an uncertainty of $\sim$0.01-0.02
mag. It may be possible to determine these corrections with an uncertainty 
of $\sim$0.01 mag, and such efforts are in progress (J. Marriner, priv. comm.). 
 
The overall ``gray'' flux scale calibration error, $\Delta_r$, is 
determined by the accuracy of the absolute flux calibration of fundamental 
standard BD+17$^\circ$4708 (Fukugita et al. 1996), the accuracy of tying the 
primary standard star network
to BD+17$^\circ$4708, the accuracy of transfering the primary standard 
star network to the secondary standard star network, and the accuracy 
of the calibration of the survey imaging data using the secondary standard 
star network. Given these numerous sources of error, it seems unlikely 
that $\Delta_r < 0.02$ mag. On the other hand, formal analysis of all 
the error contributions, as well as a direct comparison to HST observations
of hot white dwarfs (Eisenstein et al. 2006), suggest that $\Delta_r$ does 
not exceed 0.05 mag. Note, however, that all these uncertainties in the 
definition and transfer of the standard star network become moot if one 
accepts that
\begin{enumerate}
\item
$\Delta_r$ does not need to be known exquisitely well for most scientific
applications. Even if it does, this is just {\it a single number} that 
modifies the cataloged photometry for all the sources and all the bands 
in the same fashion. 
\item 
Uncertainties in the determination of $\Delta_{mr}$ are of the order 0.01 mag. 
\item
$\delta_m({\rm RA,Dec})$ can be constrained, or corrected for, at 
the 0.01 mag level. 
\end{enumerate}
In other words, the band-to-band calibration can be fixed by adopting 
$\Delta_{mr}$, determining and correcting for $\delta_m({\rm RA,Dec})$
guarantees internal consistency, and the only remaining relatively free 
parameter is $\Delta_{r}$. Such a system is then {\it no longer defined by a set
of celestial standards but rather by the functions $\phi_b$.} The catalog 
presented here is one realization of such a photometric system.

\subsubsection{        Determination of  $\delta_{m}({\rm RA,Dec})$        }

We now proceed to describe methods for constraining $\delta_{m}({\rm RA,Dec})$.
The region covered by the SDSS Stripe 82 is an elongated rectangle with
an aspect ratio of 1:50, and with the long side parallel to the Celestial Equator. 
Because of this large aspect ratio, and because different effects contribute 
to the RA and Dec dependences of $\delta_{m}$, we assume that it can be 
expressed as a sum of independent functions of RA or Dec, respectively
\eq{
 \delta_{m}({\rm RA,Dec}) = \delta_{m}^{ff}({\rm Dec}) + \delta_{m}^{ext}({\rm RA}),
}
with $\left\langle\delta_{m}^{ff}({\rm Dec})\right\rangle_{\rm RA}=0$ and 
$\left\langle\delta_{m}^{ext}({\rm RA})\right\rangle_{\rm Dec}=0$,
where $\left\langle\delta\right\rangle_x$ denotes the average of $\delta$ over 
direction $x$.

The first term, $\delta_{m}^{ff}({\rm Dec})$, is dominated by the errors in
the flatfield vectors (for drift scanning, flatfield corrections are one-dimensional). 
The flatfield determination for SDSS was difficult due to scattered 
light\footnote{For details please
see http://www.sdss.org/dr5/algorithms/flatfield.html and Stoughton et al.
(2002).} and there are probably systematic errors in the stellar photometry at 
the 0.01 mag level in the $griz$ bands and the 0.02 mag level in the $u$ band 
(perhaps somewhat larger at the edge of the imaging camera), as we will
demonstrate below.
Since these systematic errors do {\it not} cancel when averaging many
observing runs (because most stars are always observed by the same CCD
and fall on roughly the same position within the CCD), $\delta_{m}^{ff}({\rm Dec})$
could be as large as $\sim0.01-0.02$ mag on spatial scales
much smaller than the chip width of 13 arcmin. 

The second term, $\delta_{mr}^{ext}({\rm RA})$, is dominated by 
unrecognized fast variations of atmospheric extinction (e.g. due to cirrus),
because for each observing run only a single zeropoint per CCD is 
determined\footnote{Although not relevant for equatorial runs discussed here, slow 
change of atmospheric extinction due to varying airmass is accounted for;
the response of the telescope and CCDs is stable at the $<$1\% level 
on single-night timescales.}. While such variations are uncorrelated for 
different runs, it is possible that they do not average out fully at the
1\% level.

\subsubsection{ The six $ugriz$ photometric systems }
\label{sixSystems}

Before we describe methods for determining $\delta_{m}({\rm RA,Dec})$,
we address the bandpass differences between the six camera columns.
The bandpasses had been measured for each CCD using a monochromator 
(M. Doi et al. 2007, in prep.) and found not to be identical, 
as shown in Figure~\ref{compareBandpasses}. These differences between
the bandpasses induce color term errors in the reported SDSS photometry
because the magnitudes of calibration stars obtained by the PT are
transformed to the 2.5m system using a single set of color 
terms\footnote{See 
http://www.sdss.org/dr5/algorithms/jeg\_photometric\_eq\_dr1.html}.
In other words, the color difference between a blue star and a red star 
depends on which camera column the stars fall on: {\it at the $\sim$1\% 
accuracy level discussed here, there are six SDSS $ugriz$ systems.}

We used the measured response curves to generate synthetic $ugriz$
photometry corresponding to six camera columns (via 
eq.\ref{ABmag}) for 175 stars from the Gunn-Stryker atlas (1983). 
The differences between the predicted magnitudes for each camera column 
and the values generated with the response curves which define the
SDSS system\footnote{Transmission curves for the SDSS 2.5m photometric 
system are available at http://www.sdss.org/dr5/instruments/imager} 
represent photometric corrections due to bandpass
differences. These corrections may be as large as 0.02 mag for
the reddest stars, but fortunately admit simple linear fits
as a function of color ($u-g$ for the $u$ band, $g-r$ for $g$ and $r$,
and $r-i$ for $i$ and $z$). The $u$ band is the only one where 
piecewise fits are required (in the range $0.7<u-g<$1.3, the $g-r$
color provides a better fit than the $u-g$ color). We have applied 
these best-fit corrections to all sources in the catalog. The median
and rms scatter for the distributions of corrections 
evaluated for all stars, and for each color and camera column 
combination, are listed in Table~1. 

The bandpass differences have the largest impact on the $i-z$ color
of red sources. The color term errors result in a rotation of the
stellar locus in the $i-z$ vs. $r-i$ color-color diagram (see the
bottom right panel in Figure~\ref{CMD1}), and we utilize
this fact to demonstrate the improvement in photometry due to
applied corrections. We use the mean position of the stellar locus
to ``predict'' the $i-z$ color from the measured $r-i$ color,
and compute the difference between predicted and measured $i-z$
colors separately for blue (0.1$<r-i<0.2$) and red (0.8$<r-i<$1.4)
stars, and in small bins of declination (cross-scan direction). 
The difference of these residuals (blue vs. red stars) effectively 
measures the locus position angle and is not sensitive to photometric 
zeropoint errors and flatfield errors (which can only induce locus 
shifts that have no effect on this test because they cancel out). 

\begin{figure}
\plotone{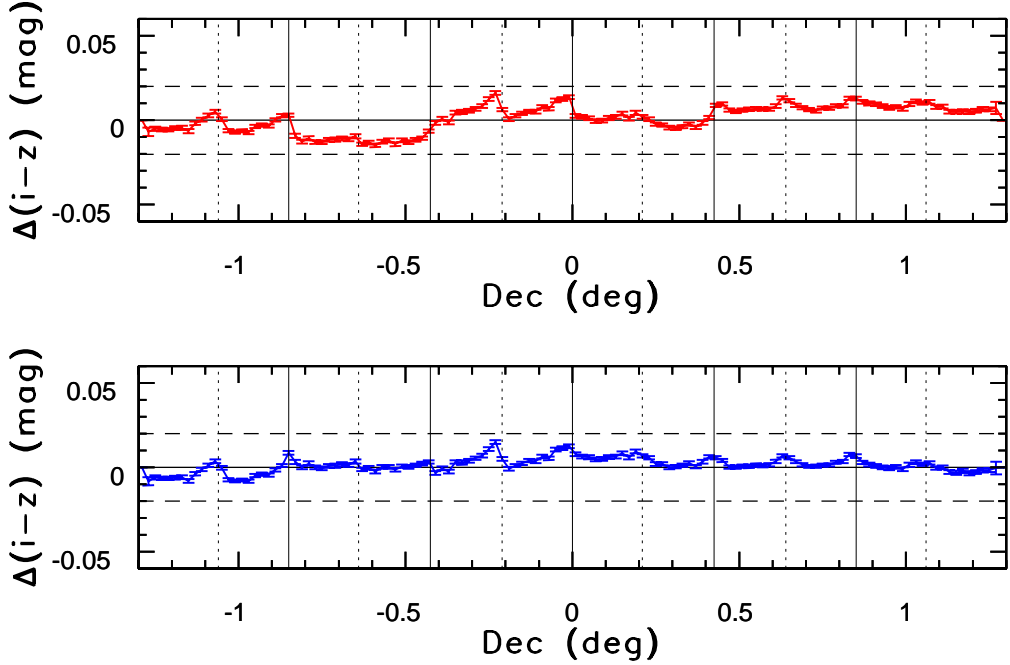}
\vspace{-4cm}
\vspace{-1.0in}
\caption{A test based on the stellar locus position angle in the $i-z$ vs.
$r-i$ color-color diagrams that demonstrates the existence of color terms
between different camera columns. The y axis shows the difference in the
$i-z$ color residuals of blue (0.1$<r-i<0.2$) and red (0.8$<r-i<$1.4)
stars, with the residuals computed as the difference between measured $i-z$
colors and those predicted using the mean stellar locus (see the bottom
right panel in Figure~\ref{CMD2}). The vertical solid lines mark the
approximate boundaries between different camera columns, with the vertical
dashed lines marking approximate boundaries between the ``north'' and
``south'' strips in a stripe (see \S\ref{overview} for definitions). The top panel 
shows results before applying corrections for different transmission curves 
(see Figure~\ref{compareBandpasses}) and the bottom panel shows results based
on corrected photometry. As evident, the residuals in the bottom panel
are much smaller, with rms scatter decreasing from 9 millimag to 3 millimag. 
\label{izResidTest}}
\end{figure}

The top panel in Figure~\ref{izResidTest} shows the median $i-z$
residuals before applying corrections for different transmission curves,
and the bottom panel shows results based on corrected photometry. 
The rms scatter decreases from 9 millimag to 3 millimag after
applying the corrections. The remaining deviations of residuals from 
zero could be due to the fact that we have not corrected for
dependence of the flatfields 
on source color (measurements of this dependence are not
available). Because this is currently the only available SDSS catalog
with photometry corrected for bandpass differences, it {\it effectively}
represents a practical definition of the SDSS photometric system (i.e. 
it provides photometry for real sources on the sky; the system is formally
defined by $\phi_b$, of course).

\subsubsection{ Other Sources of Systematic Errors }

The non-linearity of the detectors (as a function of source brightness) 
has also been measured {\it in situ} by Doi et al. and found to be 
at most a couple of percent effect over the relevant dynamic 
range\footnote{For details
see http://www.sdss.org/dr5/instruments/imager/nonlinearity.html}. 
These corrections 
are determined with a sufficient accuracy ($<$5 millimag impact on 
photometry) and are already implemented in the photometric pipeline
(i.e. before performing photometric calibration).

Similarly to the camera column-to-column bandpass differences, variations
in the wavelength dependence of atmospheric transmissivity can also
induce systematic errors that depend on source colors. Assuming
a standard atmosphere, we find using synthetic photometry for stars
from the Gunn-Stryker atlas that this effect can induce offsets of up
to $\sim$0.01 mag for the $u-g$ and $g-r$ colors when airmass
is varied by 0.3 from its fiducial value of 1.3. However, because
all stripe 82 data are obtained at the same airmass, this effect 
is not relevant for the catalog discussed here. At least in principle,
similar, and potentially larger, errors could be induced even at 
a constant airmass if the wavelength dependence of atmospheric 
transmissivity is significantly different from the assumed standard 
atmosphere. Given that such errors would probably average out,
and that there are no available measurements of the wavelength 
dependence of atmospheric transmissivity\footnote{Some handle on
the stability of the wavelength dependence of atmospheric transmissivity 
can be obtained by studying first-order extinction coefficients 
determined by the photometric calibration pipeline. They show
a cyclic variation during the year, with an rms scatter of residuals 
around the mean relation of 0.01 mag. The amplitude of the yearly
variation of the $r$ band extinction coefficient is $\sim$20\% about 
the mean value, and the wavelength dependence of the variation appears 
consistent with the addition of a gray opacity source during summer months.} 
for SDSS data considered here, we ignore this effect hereafter.

\subsection{        Determination of Flatfield Corrections }

We use two methods based on SDSS data to constrain $\delta_{m}({\rm RA,Dec})$:
a direct comparison with the secondary standard star network (\S
\ref{PTmethod}) and a method based on stellar colors (\S \ref{PCmethod}). 
Each of these methods has its advantages and disadvantages. 
We perform the tests of the final catalog quality using a method based on the 
photometric redshift relation for galaxies (\S \ref{photoz}), and also compare 
the SDSS photometry to an independent set of standards provided by Stetson 
(2000, 2005) in \S \ref{Stetson}. 

A determination of the $\delta_{m}({\rm RA,Dec})$ error from a
direct comparison with the secondary standard star network (hereafter
PT comparison) might be considered as the best method {\it a priori}. 
However, it is quite possible that the secondary standard star network itself,
off of which this catalog is calibrated, may induce 
a spatial variation of the photometric zeropoints at the 0.01 mag level 
(Smith et al. 2002). 
In addition, there are not enough stars to constrain $\delta_{m}^{ff}({\rm Dec})$ 
with sufficient spatial resolution (say, at least $\sim$100 pixels, or 
$\sim$0.01$^\circ$). For example, there are $\sim$20,000 secondary standards 
from Stripe 82 in the averaged catalog that are not saturated in the $gri$
bands in the 2.5m scans, and have PT errors smaller than 0.03 mag ($\sim$8,000
stars are useable in the $z$ band, 
and only $\sim$3,000 in the $u$ band). If these stars are binned in the 
declination
direction every 0.01$^\circ$ (250 90-pixels wide bins), $\delta_{m}^{ff}$ in each 
bin can be constrained 
to about $\sim$0.005 mag (0.01 mag in the $u$ band). This is barely
sufficient in the $gri$ bands, and cannot provide satisfactory constraints 
on the flatfielding errors in the $u$ band (where unfortunately
these errors are the largest). Similarly, $\delta_{m}^{ext}({\rm RA})$ 
can be constrained in 0.5$^\circ$ wide right ascension bins with a similar accuracy, 
but the secondary standard stars are not uniformly distributed in right ascension. 
For these reasons, we combine the PT comparison with the stellar locus method 
to determine flatfield corrections.

\subsubsection{    Color corrections from the stellar locus method       }
\label{PCmethod}
\begin{figure}
\vspace{-3cm}
\plotone{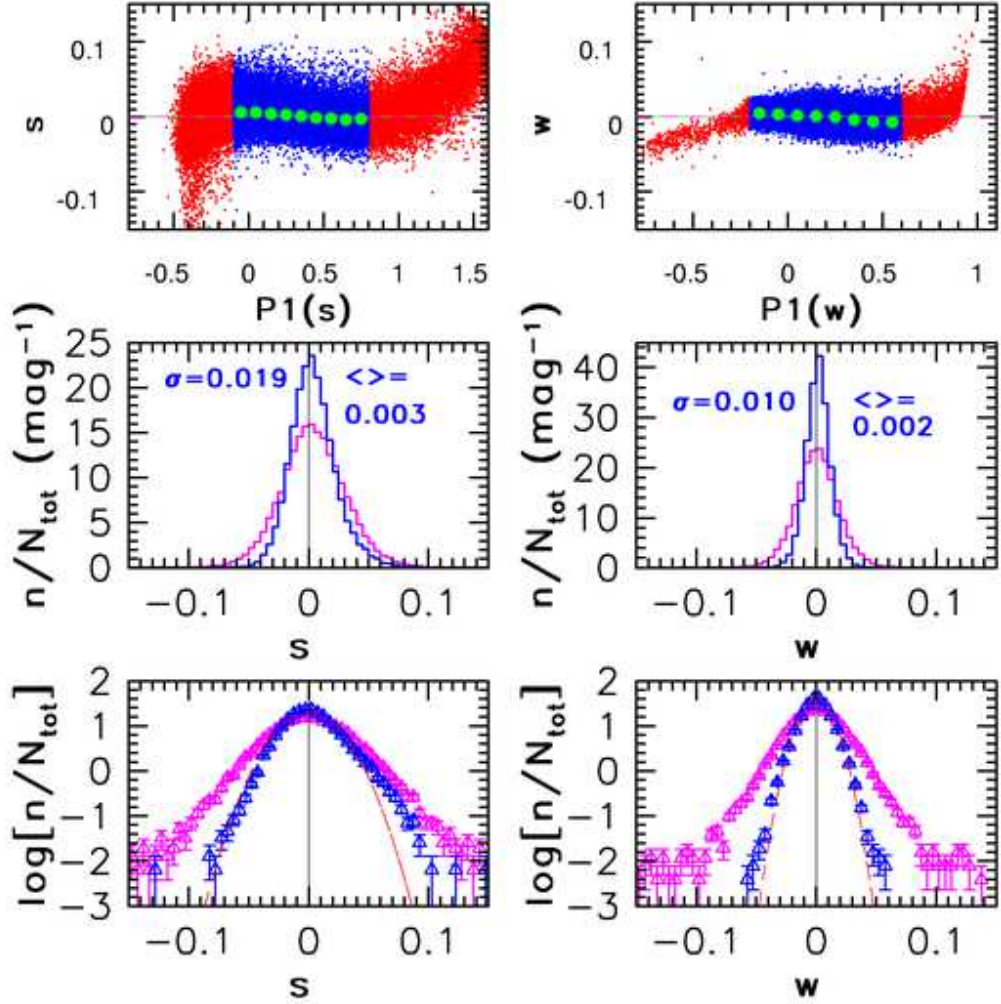}
\vspace{-1cm}
\caption{\small The top two panels show two projections of the stellar 
locus (rotated locus from the $ugr$ and $gri$ planes, see Figure~\ref{CMD2})
constructed using averaged photometry. 
The large green dots show the median values of the $s$ and $w$ principal 
colors (perpendicular to the locus at its blue edge) in bins of the 
principal color along the locus.
The middle and bottom rows show histograms for each principal color 
on linear and logarithmic scales (essentially 
the locus cross-sections). The blue
(narrow) histograms are constructed using the averaged photometry, 
and the magenta  histograms are based on single-epoch photometry.
The former are narrower than the latter due to increased
photometric accuracy. The best-fit Gaussians, with parameters listed 
in the middle row, are shown by the dashed lines in the bottom row 
(only for the averaged photometry).
\label{PC4}}
\end{figure}

The stellar distribution in color-color space at high Galactic 
latitudes\footnote{At low Galactic latitudes several effects, discussed 
below, prevent the use of this method for calibration purposes.} ($|b|>30$)
is remarkably uniform at the faint flux levels probed by SDSS,
as discussed in detail by Ivezi\'{c} et al. (2004a).  Systematic photometric 
errors, other than an overall gray error, manifest themselves as shifts in 
the position of the stellar locus that can be tracked using the four principal 
colors ($swxy$) defined by Ivezi\'{c} et al. (2004a). 
These colors are linear combinations of magnitudes,
\begin{equation}
           P_2 = A\,u + B\,g + C\,r + D\,i + E\,z + F,
\end{equation}
where $P_2 = s, w, x, y$, and measure the distance from the center of the 
locus in various two-dimensional projections of the four-dimensional stellar 
color distributions ($s$: perpendicular to the blue  part of the locus in the 
$g$-$r$ vs. $u$-$g$ plane, $w$: perpendicular to the blue part in the $r$-$i$ 
vs. $g$-$r$ plane, $x$: perpendicular to the red part, with $g-r\sim1.4$, 
in the $r$-$i$ vs. $g$-$r$ plane, and $y$: perpendicular to the locus in 
the $i$-$z$ vs. $r$-$i$ plane). The matrix of coefficients $A-F$ is listed 
in Table~2 (for more details see Ivezi\'{c} et al. 2004a). Of course, 
the measurements must be corrected for the effects of interstellar dust
extinction; we use maps provided by Schlegel, Finkbeiner \& Davis (1998, 
hereafter SFD98). The properties of two of these colors ($s$ and $w$) are 
illustrated in Figure~\ref{PC4}. 

The fact that the median principal colors are close to zero shows that the averaging 
procedure did not induce any shifts in zeropoints compared to the average 
of 291 SDSS runs which were used to define the principal colors (Ivezi\'{c} et al. 
2004a). The same conclusion
is reached by comparing the averaged photometry with the secondary
standard star network: the median photometric residuals at the so-called
crossing colors\footnote{Crossing colors are roughly the median colors of
the observed stellar population, for details see 
http://www.sdss.org/dr5/algorithms/jeg\_photometric\_eq\_dr1.html}
are 4, 6, 3, 2 and 2 millimag in the $ugriz$ bands, respectively. Yet another 
test is a direct comparison of averaged photometry with single-epoch 
photometry. Using the SDSS Data Release 5 photometry, we find that the
largest median magnitude difference between the two sets is 2 millimag 
in the $u$ band.

It is noteworthy that the widths of the principal color distributions (i.e.
the thickness of the stellar locus) constructed with averaged photometry are 
much smaller than when using single-epoch data (see the bottom four panels
in Figure~\ref{PC4}). Indeed, all four principal color distributions are 
"resolved" using this high quality photometry (see the fifth column in Table~3). 

Because the intrinsic widths of the principal color distributions are so 
small, principal colors can be used to efficiently track local calibration
problems using a small number of stars, allowing a high spatial resolution.
That is, we require that the locus not move in the 
multi-dimensional color space. In practice, the deviations of the principal
colors from 0 can be inverted, using an appropriate closure relation (see
the next section), to yield flatfield corrections (Ivezi\'{c} et al. 2004a).
With bins 0.01$^\circ$ wide in the declination direction, or 1$^\circ$ wide 
in the right ascension direction,  the flatfield corrections can be determined 
with an accuracy of 5 millimag, or better.

\subsubsection{ Gray corrections from the comparison with secondary standard star network }
\label{PTmethod}

The main advantage of the stellar locus method is that it can constrain
$\delta_{mr}$ with high spatial resolution. However, it 
is insensitive to gray errors, parametrized by $\delta_{r}$ (e.g. an 
overall gradient of photometric zeropoints in the declination direction that is
the same in
all five bands would have no effect on stellar colors). On the other
hand, the PT comparison can constrain $\delta_{r}$, but it does not 
provide enough spatial resolution to derive flatfield corrections,
especially in the $u$ and $z$ bands. Therefore, we combine these two 
methods to derive flatfield corrections $\delta_{m}^{ff}({\rm Dec})$.

\begin{figure}
\plotone{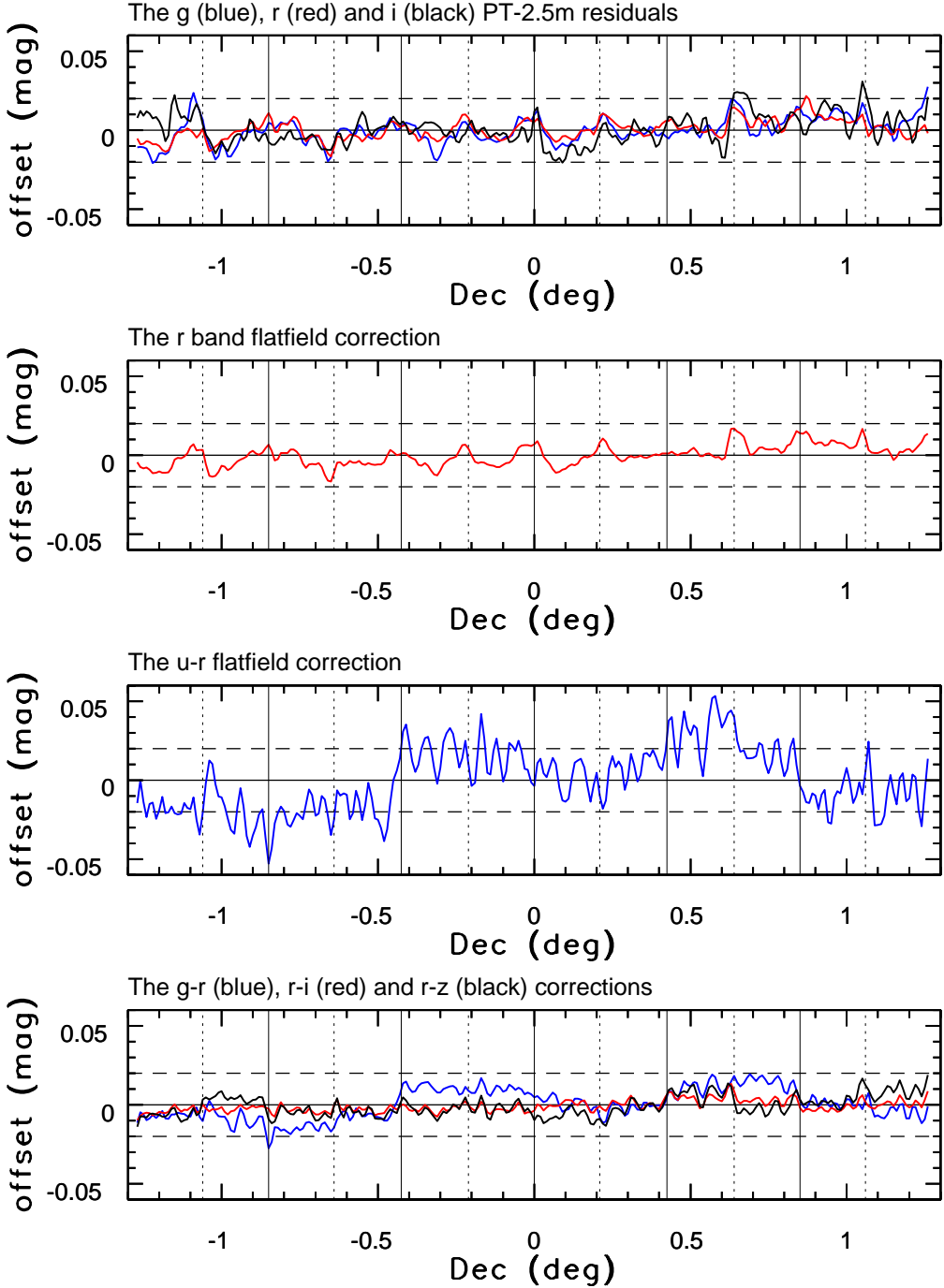}
\vskip -0.5in
\caption{The top panel shows the distribution of the residuals 
between the PT photometry and averaged magnitudes in the $gri$ 
bands. The second panel shows applied flatfield correction in the 
$r$ band, which was derived as the mean of the residuals shown in 
the top panel. The remaining two panels show the applied flatfield 
corrections in the other four bands, expressed relative to the $r$ 
band (middle: $u$, bottom: $gri$), which were derived using the 
stellar locus colors. The low-order statistics for these corrections 
are listed in Table~4.
\label{FFcorrs}}
\end{figure}

The median differences between the averaged 2.5m photometry and 
PT photometry for secondary standard stars in the $gri$ bands are shown 
in the top panel in Figure~\ref{FFcorrs}. The median differences are 
computed for 0.01$^\circ$ wide bins, and then smoothed by a triangular filter 
($y_i$ is replaced by $0.25*(y_{i-1}+2y_i+y_{i+1})$). The residuals in
all three bands display similar behavior and imply about 0.02 peak-to-peak 
variation between 
the center and edges on each CCD (resulting in about 6 millimag rms 
contribution to the overall errors), as well as an overall 0.01-0.02 mag
tilt. These systematic errors may
be due to imperfect flatfield vectors used to reduce the data, 
incorrectly determined scattered light correction (the two
are somewhat coupled in the data reduction procedures), or 
problems in PT itself (such as the PT flatfield). 

At face value, these residuals could be used to correct the averaged 
2.5m photometry in each band ($gri$) separately. However, doing so 
introduces noise in stellar principal colors of about 5 millimag (rms) and 
suggests  that the differences in photometric residuals between the 
three bands are dominated by PT measurement noise.
On the other hand, the 2.5m vs. PT residuals do 
contain information about ``gray'' errors that {\it cannot} be determined
using stellar locus. Hence, we take the {\it mean} value of the 2.5m 
vs. PT residuals in the $gri$ bands to represent the $\delta_{r}$ 
flatfield correction, and apply it to the averaged 2.5m photometry 
in the $r$ band. The applied $r$ band correction is shown in the second
panel in Figure~\ref{FFcorrs} and has an rms scatter of 7 millimag (for
250 bins), with the largest correction less than 0.02 mag. 
 
In the second step, we use the stellar locus to derive the $\delta_{mr}$ 
corrections in each band ($ugiz$). The derivation of these corrections is 
essentially identical to the procedure described by Ivezi\'{c} et al. 
(2004a). Also, together with a PT-based $\delta_{r}$ correction, 
this is essentially the same method as used to derive flatfield 
corrections for the whole SDSS survey\footnote{For details see 
http://www.sdss.org/dr5/algorithms/flatfield.html.}. In particular,
we used here the same closure relation (stellar locus method gives
four equations for five unknowns), that is based on averaged 2.5m vs. 
PT residuals in the $gri$ bands. The resulting flatfield corrections,
$\delta_{mr}$, in the $ugiz$ bands are shown in the third and
fourth panels in Figure~\ref{FFcorrs} and summarized in Table~4.

Due to low stellar counts and the strongest scattered light, the $u$ band 
correction is expected to have the largest noise 
($\sim$5-10 millimag), which is consistent with the observed behavior.
It is thus likely that some of the variation on scales of $\sim$0.01$^\circ$
is not real. On the other hand, it could be argued that systematic errors 
could actually be much larger on even smaller spatial scales, but get 
averaged out in 90 pixel wide bins. However, in addition to not having 
a reason to believe in such high spatial frequency effects (e.g. the sky 
background does not show any evidence for them), no additional scatter, 
except the expected statistical noise, is observed when the bin 
size is decreased by a factor of 4. 

Last but not least, it is important to emphasize that these corrections
are {\it not} setting photometric zeropoints, but only correcting for variations 
in response across each CCD. As discussed above, the AB photometric zeropoints, 
relative to the fiducial $r$ band, {\it are effectively set by adopting values
for} $\Delta_{mr}$.

\subsection{                     Tests of Catalog Quality                      }

\subsubsection{                     The internal tests                 }
\label{internal}
At least in principle, the same methods used to derive $\delta_{m}^{ff}({\rm Dec})$
could be used to derive $\delta_{m}^{ext}({\rm RA})$. However, in practice
this is not possible for at least two reasons: first, the right ascension distribution 
of secondary standard stars is not as uniform as their declination distribution, and
second, the assumption of the constancy of the stellar locus in color space
is invalid along the ``long'' scan direction (discussed below). For these
reasons, we only use the PT comparison and stellar locus methods to estimate the 
level of internal zeropoint variations with RA, and do {\it not} correct the 
data. In the next section, we will also use another method, based on 
galaxy colors, as an independent test of catalog integrity. 

\begin{figure}
\plotone{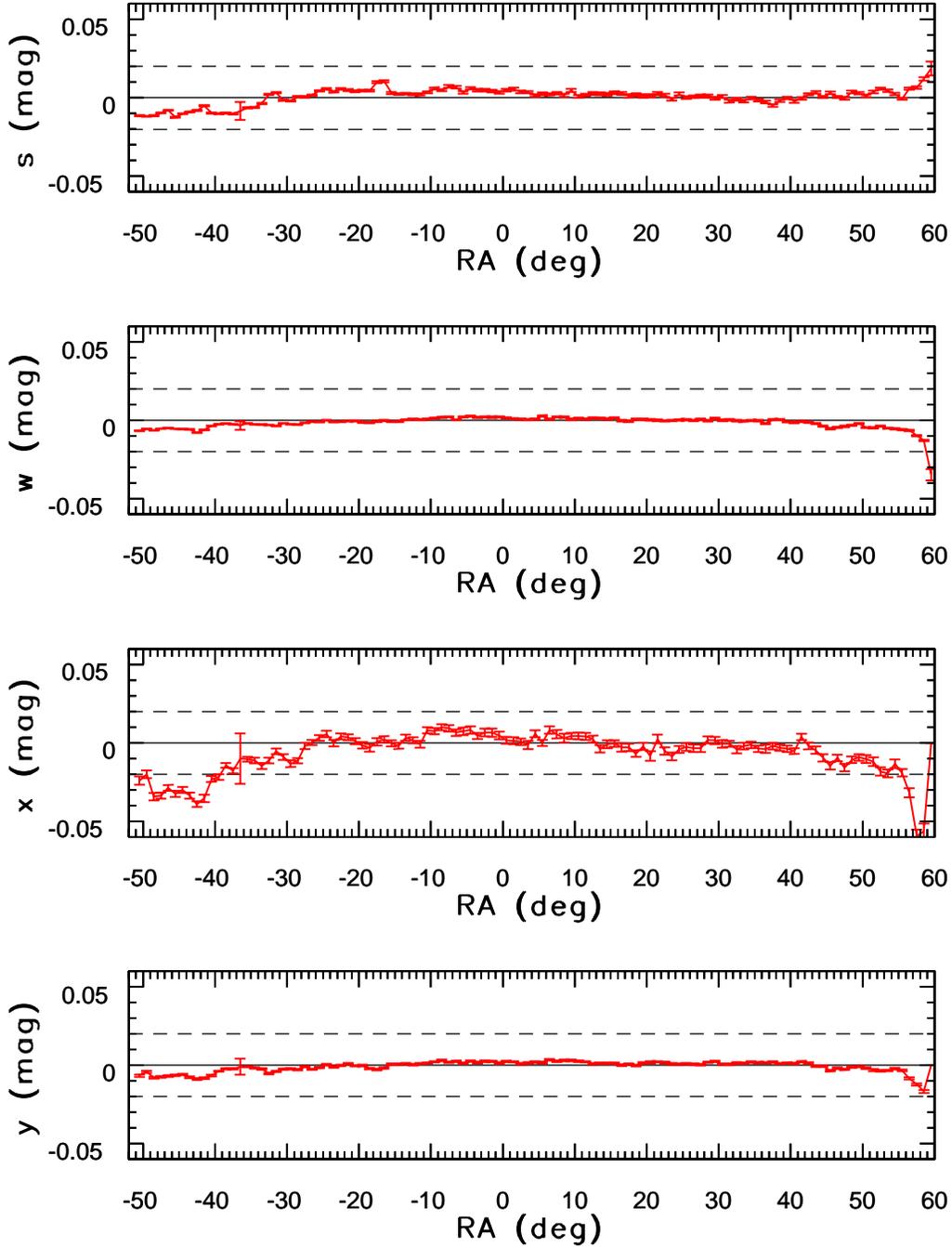}
\caption{The panels show the dependence of the position of the
stellar locus in $ugriz$ color space, as parametrized by
the median principal colors $swxy$, as a function of RA. 
Close to the edges, the median colors deviate significantly
from 0. This is caused by intrinsic changes in the stellar locus
due to stellar population variations and overestimated interstellar 
extinction corrections for red stars, rather than calibration problems.
\label{PCsRA}}
\end{figure}

\begin{figure}
\epsscale{.8}
\plotone{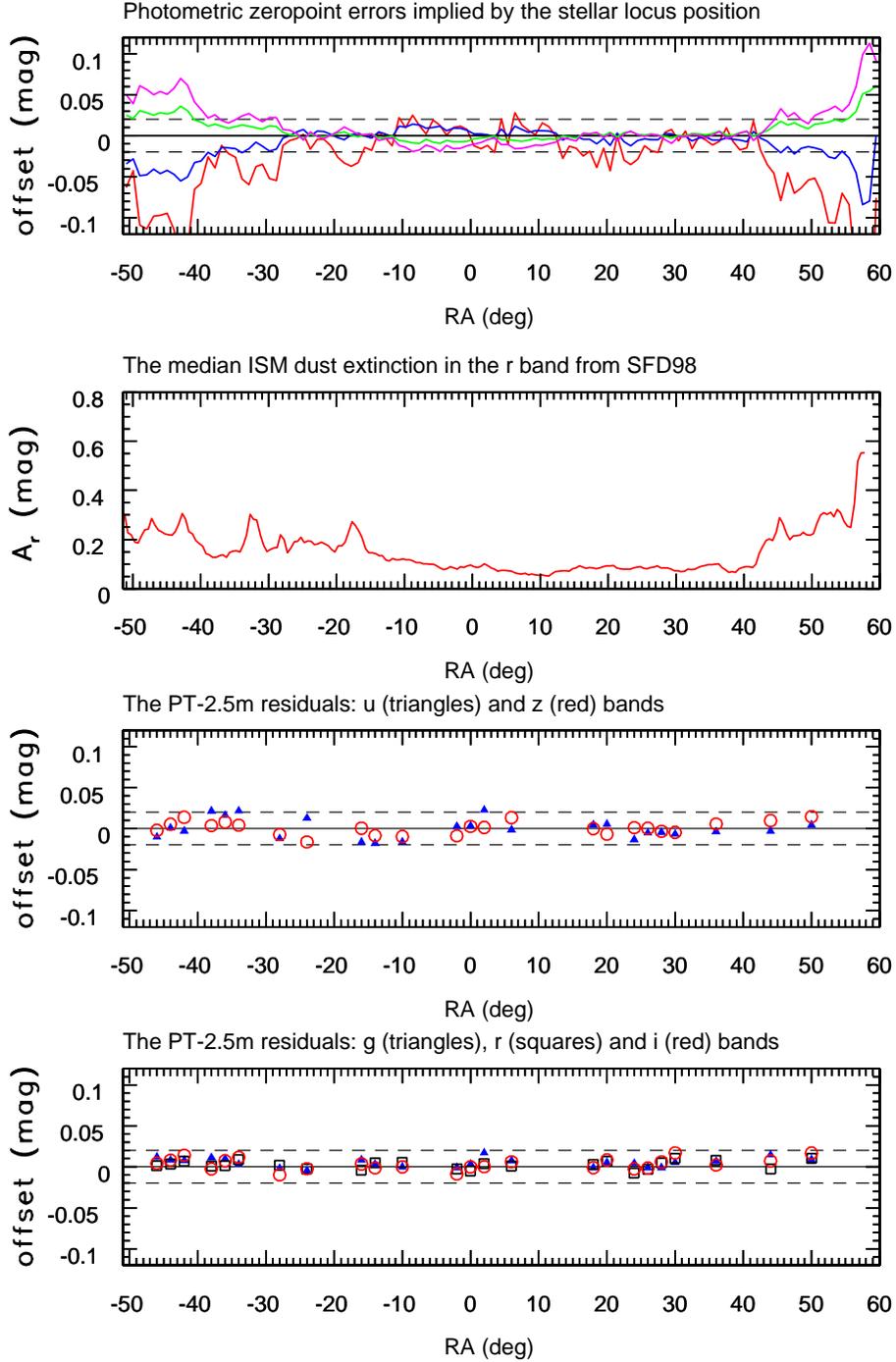}
\vspace{-13cm}
\caption{\small  The top panel shows the implied photometric
zeropoint errors based on the stellar locus method
($ugiz$ from the bottom to the top at either edge). 
While the implied errors are small for $-25 <$ RA $<40$, 
they become exceedingly large outside this range. This
is due to problems with the stellar locus method rather
than due to problems with calibration. The second panel
shows the median $r$ band extinction derived from the Schlegel,
Finkbeiner \& Davis (1998) maps. The bottom two panels
show the median residuals between the PT photometry and 
averaged magnitudes in the $uz$ (third row) and $gri$ 
bands (bottom row). The low-order statistics for these 
residuals are listed in Table~5.
\label{P8}}
\end{figure}

Figure~\ref{PCsRA} shows the median principal colors in bins of right ascension. 
As evident, the principal colors are close to zero for the right ascension
in the range $-25^\circ <$ RA $<40^\circ$,
but outside this range deviate significantly from zero. This does not 
necessarily indicate problems with photometric calibration, because
the stellar locus method is expected to fail at low Galactic latitudes
due to several reasons. First, the mean metallicity of stars increases
at low Galactic latitudes and this change may affect the $s$ and $w$ colors. 
Second, at low latitudes red dwarfs are not behind the entire dust screen 
measured by the SFD98 maps (see Juri\'{c} et al. 2006 for a discussion of this 
point), and thus the $x$ color will be biased blue (i.e. the colors of red 
dwarfs are over-corrected for the ISM reddening). And third, at low latitudes the dust column 
increases fast (see Figure~\ref{P8}) and even small errors in the {\it assumed} 
wavelength dependence of the dust extinction, or the extinction itself
as given by the SFD98 maps, will have noticeable effects on principal colors. 
For these reasons, it seems plausible that the deviations seen in 
Figure~\ref{PCsRA} are {\it not} dominated by zeropoint errors. 

This conclusion is supported by the direct comparison of the averaged and PT 
photometry (Figure~\ref{P8}). For example, the largest median photometric 
residual between the averaged catalog and PT observations in the $u$ band 
is $\sim$0.02 mag (see Table~5), which is much smaller than the 0.1 mag 
discrepancy implied by the stellar locus method. 

Table~5 shows that the rms scatter of median photometric residuals (evaluated
in 2$^\circ$ wide bins in the right ascension direction) between the averaged catalog and PT observations 
is $<$0.01 mag in all five bands. Some of that scatter 
must come from the PT data itself, and thus the true scatter of photometric zeropoints
in the averaged catalog is probably even smaller than that listed in Table~5. 
In addition, Table~5 shows that the averaged catalog and PT measurements are on the 
same system to within a few millimags (using the recommended photometric 
transformations between the two telescopes listed at the SDSS website, see 
Section~\ref{photomCalib}).

We have also compared the catalog presented here to photometric reductions
described by Padmanabhan et al. (2007). As discussed in the Introduction,
they determined flatfields by minimizing errors in relative photometry of 
multiply observed stars over the whole survey region. Hence, this comparison
is an essentially independent test of flatfield corrections derived here
despite the fact that both catalogs are based on the same observations. 
We bin the photometric differences between the catalogs in 0.01$^\circ$ wide 
Dec bins and compute the median residual for each bin and band. The median 
value of these medians represent zeropoint offsets in each band and are equal to 
-7, 2, -1, 2 and 5 millimag in the $ugriz$ bands, respectively. The rms 
scatter of the median residuals reflects systematic errors due to flatfield 
errors and we measure 27, 6, 5, 6, and 8 millimag in the $ugriz$ bands, 
respectively. Except in the $u$ band, these values indicate that systematic
flatfield errors are very small. In the $u$ band, Padmanabhan et al. (2007) expect 
errors
of about 0.01 mag, and the distribution width of the $s$ color implies about 
0.01 mag for the catalog discussed here, predicting about 14 millimag instead 
of the measured 27 millimag. It is plausible
that the $u$ band photometry may contain systematic errors unrecognized 
by any of the methods discussed here.

\subsubsection{ The tests of catalog quality based on galaxy colors}
\label{photoz}
The color distribution of galaxies is bimodal (Strateva et al. 2001,
Yasuda et al. 2001, Baldry et al. 2003). Red galaxies have an especially
tight color-redshift relation (Eisenstein et al. 2001), with an rms of 0.12 mag 
for the $u-g$ color, 0.05 mag for $g-r$, and 0.03 mag for the $r-i$ and 
$i-z$ colors (using model magnitudes).
Deviations from the mean relations can thus be used to track local
calibration problems. Of course, since this is a color-based method,
it can only constrain $\delta_{mr}$, and, because red galaxies are 
faint in the $u$ band, cannot achieve high spatial resolution in this
band. Nevertheless, it is a useful
addition to the stellar locus method because it is independent of 
the Milky Way structure and secondary star network (although it is
sensitive to errors in the ISM dust extinction correction). 

We select 19,377 red galaxies with SDSS spectra from the redshift range 
0.02--0.36 using an empirical condition
\eq{
      0 < (g-r) - 0.6 - 2.75 \times {\rm redshift} < 0.3,
}
and determine their median colors as a function of redshift using
0.01 wide redshift bins. The residuals from the median color--redshift
relation are then binned
by declination to constrain $\delta_{m}^{ff}$ and by right ascension to constrain $\delta_{m}^{ext}$.
The rms for color residuals and the widths of distributions of residuals 
normalized by statistical noise (based on quoted photometric errors) are 
listed in Table~6. 

\begin{figure}
\plotone{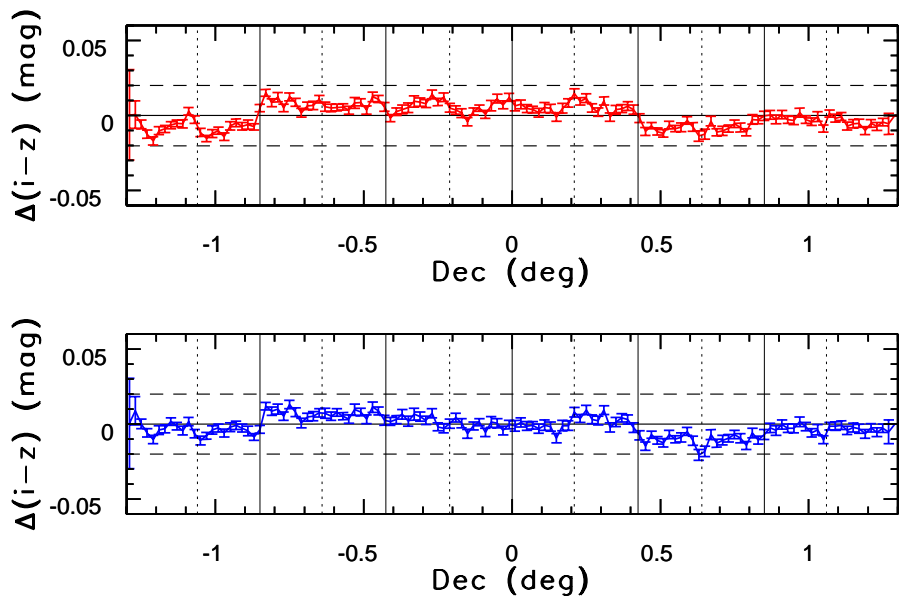}
\vspace{-2cm}
\caption{The top panel shows the dependence of the median $i-z$ color 
residuals (with respect to a mean color-redshift relation) for red galaxies 
as a function of Dec. The vertical solid lines mark the approximate boundaries 
between different camera columns, with the vertical dashed lines marking 
approximate boundaries between the ``north'' and ``south'' strips in a 
stripe. The small rms scatter of only 9 millimag demonstrates that flatfield
corrections based on the stellar locus position in color space are also 
applicable for galaxies. The bottom panel shows the difference between the 
values shown in the top panel and the curve shown in the bottom panel in 
Figure~\ref{izResidTest}. The rms scatter for the residuals in the 
bottom panel is 6 millimag.} 
\label{izGal}
\end{figure}

The residuals binned in the declination direction are generally small and
consistent with statistical noise. The largest deviations from 0 are 
seen for the $i-z$ color, with an rms of 9 millimag and maximum deviation
of 17 millimag (see top panel in Figure~\ref{izGal}). Although the rms 
is fairly small, it is a factor of 2.9 larger than the expected noise.
The shape of the $i-z$ residuals is similar to the $i-z$ residuals for 
stars discussed in \S~\ref{sixSystems} and shown in the bottom panel in 
Figure~\ref{izResidTest}. The rms scatter for the difference
between the stellar and galaxy $i-z$ residuals, shown in the bottom 
panel in Figure~\ref{izGal}, is 6 millimag. Since the rms for galaxy 
residuals is larger (9 millimag), it is plausible that the same systematic effect 
dominates the remaining photometric errors for both stars and galaxies.
While it is not clear what the cause of this (small) effect is, 
a plausible explanation is the dependence of flatfields on source 
color\footnote{It is fair to ask whether the applied flatfield 
corrections, derived from stellar colors, are actually appropriate for galaxies. They are since the 
$i-z$ color residuals for galaxies without any flatfield corrections 
are about twice as large than those shown in the top panel in 
Figure~\ref{izGal}.}. 

\begin{figure}
\epsscale{.7}
\plotone{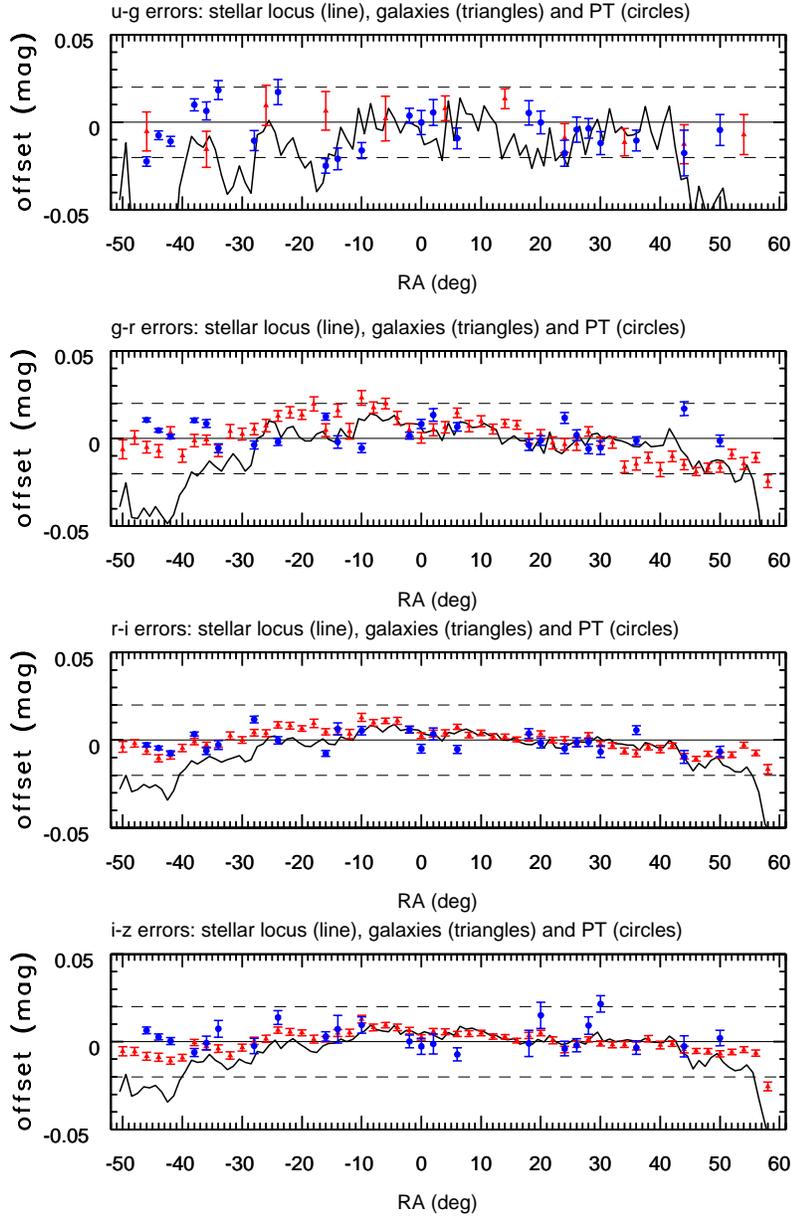}
\epsscale{.9}
\vspace{-8cm}
\vspace{-1.0in}
\caption{The comparison of systematic color errors implied
by different methods. Note that the errors implied by
the stellar locus method (line) become very large outside
the $-25^\circ < \alpha_{J2000} < 40^\circ$ range. As galaxy colors 
(triangles) and a direct comparison with SDSS secondary standard star 
network (circles) suggest, this is due to problems with the 
stellar locus method rather than due to problems with 
calibration. 
\label{P10}}
\end{figure}

The results for binning in the right ascension direction are shown in Figure~\ref{P10}, where
we compare different methods. The rms for implied
color errors (with 2$^\circ$ wide bins) from galaxies is 0.006 mag for the $r-i$ 
and $i-z$ colors, 0.012 mag for the $g-r$ color, and 0.018 mag 
for the $u-g$ color. The overall behavior of red galaxy
color residuals agrees better with the PT method, than with
color errors implied by the stellar locus method. In particular,
the large errors outside the $-25^\circ <$ RA $<40^\circ$ range
implied by the latter method are not consistent with red 
galaxy color residuals. On the other hand, both red galaxy 
colors and stellar locus seem to show a trend that the colors
are {\it redder} around RA=$-$10$^\circ$ than around RA=40$^\circ$.
The amplitude of this effect varies from about 0.02 mag for 
the $g-r$ color to about 0.01 mag for the $i-z$ color, while 
the upper limit on such a slope implied by the PT comparison 
is $<$0.01 mag.

It is not clear what the cause of this discrepancy is.
The obvious culprit is the correction for interstellar dust
extinction, but the implied deviation is too large to be
explained by any plausible errors in the SFD98 maps. 
As shown in Figure~\ref{P8}, the median extinction in 
the $r$ band for the $-10 <$ RA $<40$ range is below 0.1 mag, 
and the resulting median correction for, e.g., the $g-r$ color 
is below 0.04 mag. Hence, to induce a 0.02 mag trend in 
the $g-r$ color, the SFD value for the $r$ band extinction, $A_r$, 
would have to be in error by 0.05 mag (the difference between 
the values provided for RA=$-$10$^\circ$ and RA=40$^\circ$). This implies
relative errors  for the SFD map in the range from 50\% 
(if $A_r$ at RA=$-$10$^\circ$ is underestimated) to 100\% 
(if $A_r$ at RA=40$^\circ$ is overestimated), which seems unlikely 
(although not impossible). 

We conclude that the PT comparison provides a good estimate 
of the remaining zeropoint errors in the catalog, as listed
in Table~5, but caution that we do not understand the above
systematic behavior of stellar and galaxy colors, and that
only the PT constrains possible gray errors. In the next Section, 
we discuss a comparison to an external dataset, that supports 
this conclusion.

\subsubsection{An external test of catalog quality based on Stetson's standards}
\label{Stetson}

The only large external dataset with sufficient overlap, depth and accuracy to 
test the quality of the Stripe 82 catalog is that provided by Stetson (2000, 2005). 
Stetson's catalog lists photometry in the $BVRI$ bands (Stetson's photometry 
is tied to Landolt's standards) for $\sim$1,200 stars 
in common (most have $V<19.5$). We synthesize the $BVRI$ 
photometry from SDSS $gri$ measurements using photometric transformations 
of the following form
\begin{equation}
\label{BVRItrans}
  m_{\rm Stetson} - m_{\rm SDSS} = A\,c^3 + B\,c^2 + C\,c + D,
\end{equation}
where $m_{\rm Stetson}=(BVRI)$ and $m_{\rm SDSS}=(g,g,r,i)$, respectively, and the color $c$ is  
measured by SDSS ($g-r$ for the $B$ and $V$ transformations, and $r-i$ for 
the $R$ and $I$ transformations). The measurements are {\it not} corrected for 
the ISM reddening. 
Traditionally, such transformations are assumed to be linear in 
color\footnote{For various photometric transformations between the SDSS
and other systems, see Abazajian et al. (2005) and 
http://www.sdss.org/dr5/algorithms/sdssUBVRITransform.html.}.
We use higher-order terms in eq.~\ref{BVRItrans} because at the 1-2\% level 
there are easily detectable deviations from linearity for all color choices
(for details and plots see Ivezi\'{c} et al. 2007).  
The best-fit coefficients for the transformation of SDSS $gri$ measurements 
to the $BVRI$ system\footnote{
The same transformations can be readily used to transform measurements
in the $BVRI$ system to the corresponding $gri$ values because
$B-V=f(g-r)$ and $R-I=f(r-i)$ are monotonic functions.}, and 
low-order statistics for the 
$m_{\rm Stetson} - m_{\rm SDSS}$ difference distribution\footnote{Note 
that these transformations are valid {\it only} for main 
sequence stars with colors in the range $g-r>0.2$ and $r-i<1.5$ 
(roughly, 0.3$< B-V <$1.6). Extrapolation outside this range may result in 
large errors!} are listed in Table~7.
We find no trends as a function of magnitude at the $<0.005$ mag level.
With the listed transformations, {\it the SDSS catalog described here could also 
be used to calibrate the data to the BVRI system with a negligible loss 
of accuracy due to transformations between the two systems}.

The $BVRI$ photometry from Stetson and that synthesized 
from SDSS agree at the level of 0.02 mag (rms scatter for the magnitude 
differences  of {\it individual} stars; note that the systems are tied to 
each other to within a few millimags by the transformations listed in Table~7). 
This scatter is consistent with the claimed accuracy
of both catalogs (the magnitude differences normalized by the 
implied error bars are well described by Gaussians with widths in
the range 0.7--0.8). This small scatter allows us to test for the 
spatial variation of zeropoints between the two datasets, despite
the relatively small number of stars in common.

Stars in common are found in four isolated regions
that coincide with historical and well-known
Kapteyn Selected Areas 113, 92, 95, and 113. 
We determine the zeropoint offsets between the SDSS 
and Stetson's photometry for each region separately
by synthesizing $BVRI$ magnitudes from SDSS $gri$ photometry,
and comparing them to Stetson's measurements.
The implied zeropoint errors (which, of course, 
can be due to either the SDSS or the Stetson dataset, or both) are listed
in Table~8. For regions 1-3 the implied errors are only a few
millimags (except for the $B-g$ color in region 1). The discrepancies
are much larger for the three red colors in region 4. A comparison
with the results of internal SDSS tests described in \S \ref{internal}
and \ref{photoz} suggests that these discrepancies 
are more likely due to zeropoint offsets in Stetson's photometry for 
this particular region, than to problems with SDSS photometry. 
We contacted P. Stetson who confirmed that his 
observing logs were consistent with this conclusion. Only a small 
fraction of stars from Stetson's list are found in this region. 

Given the results presented in this Section, we conclude\footnote{Here we 
assumed that it is a priori unlikely that the SDSS and Stetson's zeropoint
errors are spatially correlated.} that the rms for the 
spatial variation of zeropoints in the SDSS Stripe 82 
catalog is below 0.01 mag in the $gri$ bands.


\section{The Utility of the SDSS Stripe 82 Standard Star Catalog}

As examples of the use of the standard star catalog, we discuss 
calibration of data obtained in non-photometric conditions, and 
a detailed and robust measurement of the morphology of the stellar 
locus in color-color space.

\subsection{         Calibration of Non-photometric Data         }

The existence of a technique to photometrically calibrate non-photometric 
data would greatly increase the efficiency of telescopes. As one particular
example of how our catalog can support a large project, consider the 
SDSS-II Supernova survey (Sako et al. 2005). This survey aims to obtain repeat 
images of stripe 82 with a 
sufficient cadence to enable discovery of new type Ia supernovae. This requirement
sometimes results in observations obtained through clouds with
several magnitudes of extinction. In such highly non-photometric
conditions the standard photometric calibration described in section 
\ref{photomCalib} fails because the fields with standard stars are 
too sparsely distributed to be able to resolve fast variations in 
cloud extinction.

\subsubsection{ A method to track fast cloud extinction variations }
\label{cloudCalibration}

Due to its high stellar density, the standard star catalog described in this paper
can be used for calibration of data obtained in grossly non-photometric
conditions. The typical number of calibration stars in each SDSS field (9$\times$13 arcmin$^2$)
at high Galactic latitudes is 10-15 in the $u$ band, 40-50 in the $gri$ bands, 
and 30-40 in the $z$ band. Based on tests of several non-photometric SDSS-II
runs, it was found that the cloud extinction variations can be tracked with a sufficient
temporal resolution ($\sim$3 sec) to obtain photometric zeropoint accuracy 
comparable to that characteristic for photometric nights (1-2\% in $gri$
and 2-3\% in $u$ and $z$; Ivezi\'{c} et al. 2004a). 

The calibration is done in two steps. First, the implied zeropoints (whose
variation is dominated by cloud extinction), $zp$, defined by 
\eq{
           m_{calibrated} =  -2.5*\log(counts) + zp,
}
are computed for each star and median filtered in time using a window with 5
stars in order to avoid outliers. Note that we assume that clouds are gray
and do not allow for color terms, an assumption which is justified 
a posteriori (see \S \ref{cloudcolor}).
In the second step, zeropoints are evaluated 
for each 2048 pixel wide (cross-scan
direction) and 100 pixel long (in-scan direction) image segment, hereafter 
called a calibration patch (not to be confused with secondary star patches
discussed in \S \ref{photomCalib}). That is, a calibration patch is a $\sim$9 
arcmin$^2$ large rectangle with an aspect ratio 1:20, and the zeropoints are 
evaluated every 2.6 seconds of time (but note that the variations are
smoothed out by the 54 sec long exposure time). 
 
\begin{figure}
\epsscale{1.2}
\plotone{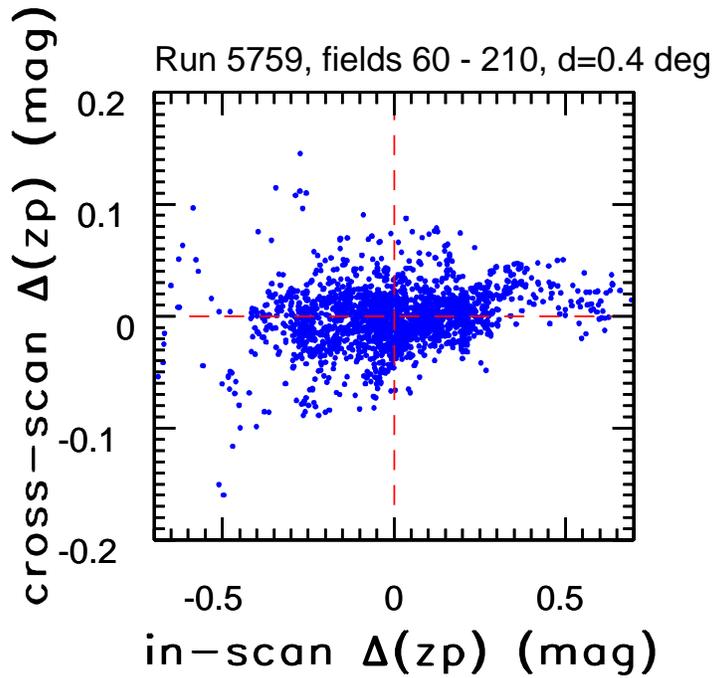}
\vskip -5in
\caption{The comparison of cloud extinction gradients in the 
in-scan (RA, horizontal axis) and cross-scan (Dec, vertical axis)
directions for SDSS run 5759, on a spatial scale of $\sim$0.4$^\circ$
(each point is derived using zeropoints from three calibration patches). 
Note the different axis scales. For this particular run, the distribution width 
is 9.6 times larger for the in-scan than for the cross-scan direction.
This is a consequence of cloud motion relative to the boresight,
and the drift-scanning technique.
\label{ZPgrad}}
\end{figure}

The patch is much narrower in the in-scan direction because tests have shown
that zeropoint gradients across a field are much larger, by a factor 10-50, 
in this direction (see Figure~\ref{ZPgrad}). Consider three 
stars, star A, a star B that is, say, 25 arcmin (the column-to-column separation)
away from the star A in the scan direction, and a star C that is 25 arcmin away 
from the star A in the cross-scan direction. Stars A and C are observed at the same 
time and the difference in their implied zeropoints measures the structure
function of cloud opacity on a 25 arcmin spatial scale. This is true
irrespective of the cloud motion relative to the boresight. Here, the
structure function of cloud opacity (SF hereafter) is defined as the rms width 
of the distribution of zeropoint differences evaluated for pairs
of points separated by some distance.

On the other hand, stars A and B are observed at times that differ by 1.7 minutes. 
If the component of the cloud angular velocity on the sky relative to the boresight and
parallel to the scanning direction is $\omega$ $^\circ$ min$^{-1}$, the zeropoint difference for 
stars A and B samples the cloud structure on spatial scales of 25 $\omega/\omega_s$ arcmin, 
where $\omega_s$=0.25 $^\circ$ min$^{-1}$ is the sidereal scanning rate (here for simplicity we 
assumed $\omega \gg \omega_s$, which is
supported by the data). The observed behavior of zeropoints, such as that shown 
in Figure~\ref{ZPgrad}, implies wind velocity in the range\footnote{This range is 
equivalent to angular speeds of up to a half of the Moon's diameter per second. 
The plausibility of this wind velocity range was verified in extensive visual 
observations of the full Moon during frequent grossly non-photometric nights in 
Seattle.} 
$\omega$=$3-15$ $^\circ$ min$^{-1}$, or $\omega/\omega_s \sim 12-60$. Hence, drift 
scanning has the unfortunate
property that the motion of an inhomogeneous extinction screen with a speed
much larger than the sidereal scanning rate greatly magnifies the effective 
zeropoint variations in the scan direction. 

The zeropoints for each calibration patch are computed by taking all the stars
from the patch, or finding the closest three stars for sparsely populated patches, 
and adopting the median value of their zeropoints. This is certainly 
not the only, nor perhaps the ideal approach to calibrate patches, but we found that it
works well in practice. The zeropoint error is evaluated from the
root-mean-square scatter of $zp$ evaluated for each calibration star, divided by 
the square root of the number of stars. We now discuss the performance of this method.

\subsubsection{ Performance and Quality Tests }

\begin{figure}
\epsscale{.8}
\plotone{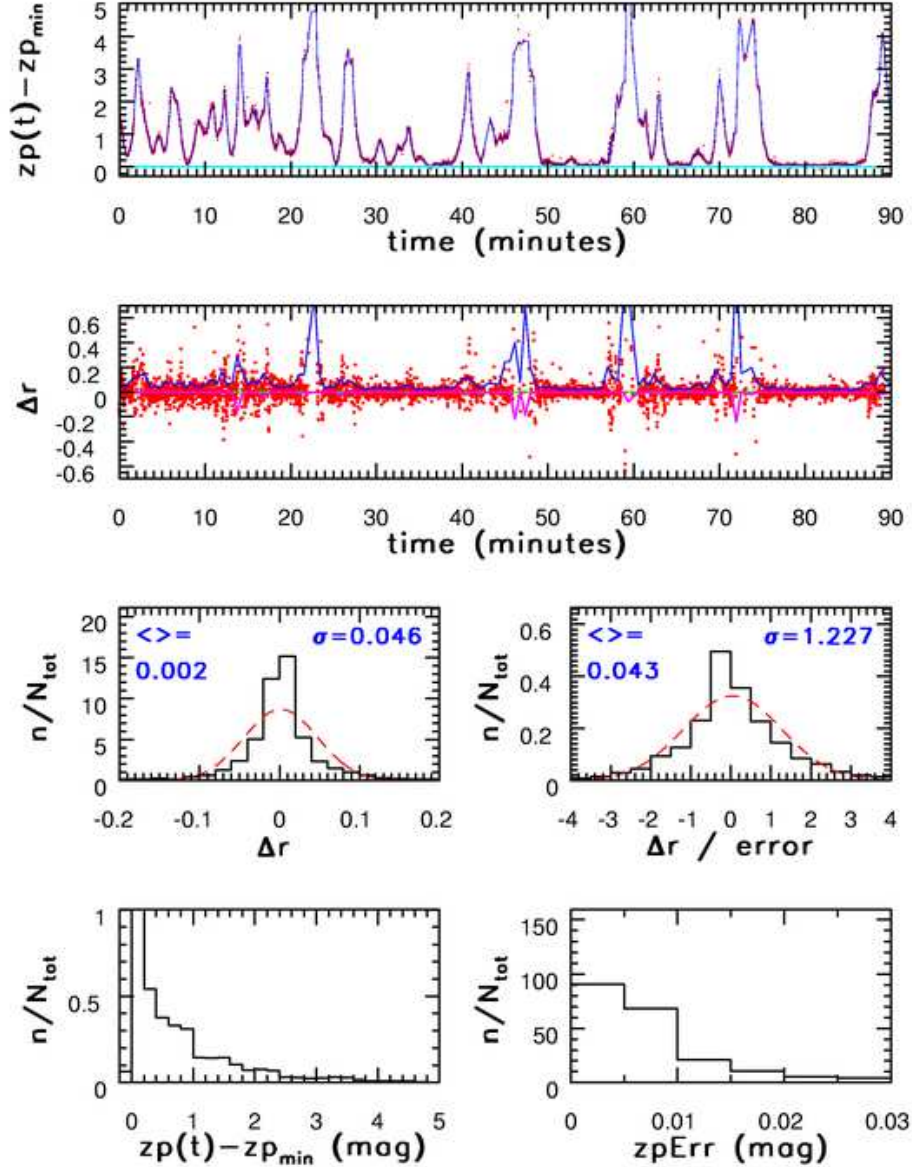}
\epsscale{.9}
\caption{\footnotesize 
The top panel summarizes the behavior of cloud extinction in the 
$r$ band over 1.5 hours during SDSS-II run 5646. Individual calibration 
stars are shown by dots, and the adopted zeropoint is shown by the line. 
The calibration residuals for each star are shown by dots in the second 
panel. The root-mean-square scatter for these residuals evaluated
for each field is shown by the line. The distribution of the residuals 
is shown in the left panel in the third row as the solid line. The median 
and equivalent Gaussian $\sigma$ evaluated from the inter-quartile range 
are also shown in the panel, as well as a Gaussian corresponding to these 
parameters (dashed line). The right panel is analogous, except that the 
residuals are normalized by the expected errors. The distribution of implied 
cloud extinction is shown in the bottom left panel, and the distribution of 
standard errors for the adopted photometric zeropoints (computed from the
rms width of the distribution of residuals) is shown in the bottom right panel
(a few points, about 4 out of 1800 calibration patches, are off scale). 
\label{dmResid1}}
\end{figure}

\begin{figure}
\plotone{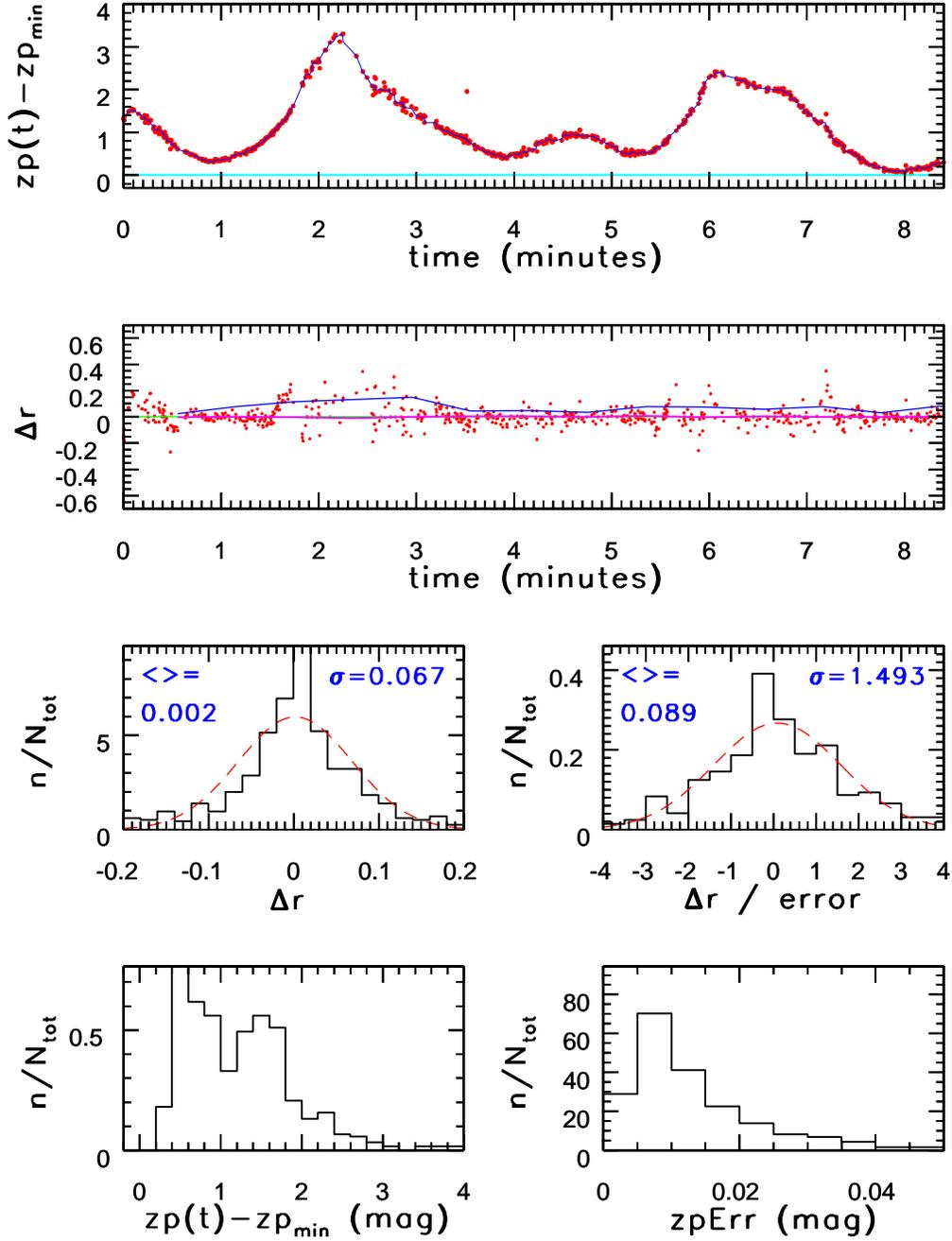}
\caption{Analogous to Fig.~\ref{dmResid1}, except that only $\sim$8
minutes of data with large cloud extinction is shown. Note 
that the changes in cloud extinction are resolved down to time scales 
well below 1 minute.
\label{dmResid2}}
\end{figure}

\begin{figure}
\plotone{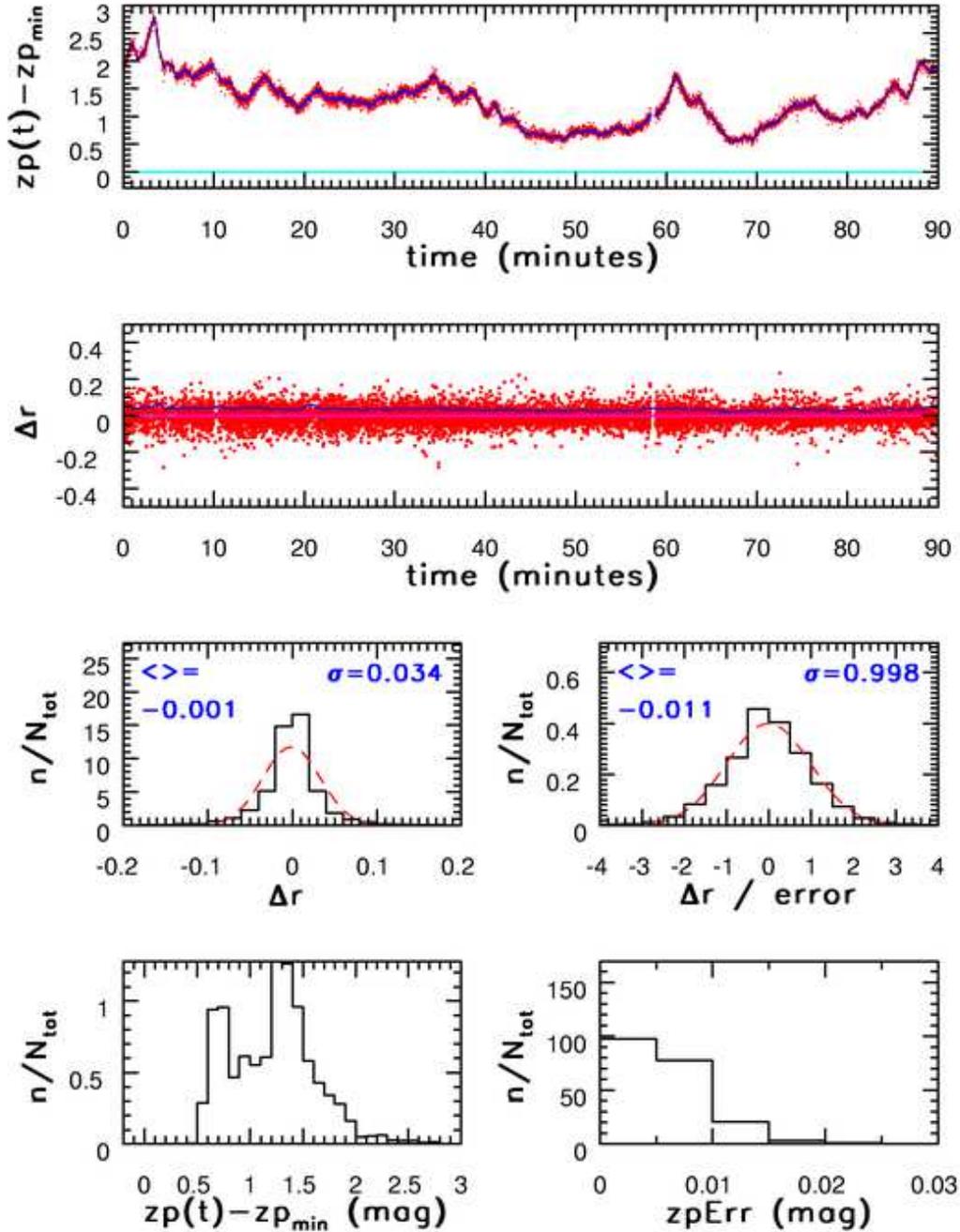}
\caption{Analogous to Fig.~\ref{dmResid1}, except that 1.5 hours 
of data from run 5759, which had somewhat thinner and much more 
stable cloud cover, is shown. Note that the median photometric zeropoint 
error is below 0.01 mag, although the median cloud extinction is larger 
than 1 mag. 
\label{dmResid2Y}}
\end{figure}

The top panel in Figure~\ref{dmResid1} summarizes the behavior of cloud
extinction in the $r$ band, as measured by the zeropoint $zp$ discussed
above, for an SDSS-II SN run (5646) obtained in strongly
non-photometric conditions. Although the cloud extinction during the first 
90 minutes (corresponding to 150 SDSS fields) varies between 0 and $\sim$6
mag, it is possible to robustly calibrate these data. Figure~\ref{dmResid2}
zooms in on a 8 minute stretch of the same data where the cloud extinction
varies between 0 and $\sim$3 mag, with changes as fast as 0.05 mag s$^{-1}$
(almost 2 mag per SDSS field!). 
As shown in the figure (middle left panel), the residuals have a distribution 
width of only 0.07 mag. The middle right panels in Figures~\ref{dmResid1} and
\ref{dmResid2} demonstrate that most of this scatter is contributed by random photometric
errors (i.e. errors in extracted source counts), rather than by calibration 
errors (large cloud extinction results in a smaller number of calibration stars, 
as well as in a lower SNR for those calibration stars that are detected). 
Even with such a large and rapidly varying cloud extinction, the zeropoint errors 
are smaller than 0.05 mag, with a median value of less than 0.02 mag. An example
of a run with somewhat thinner and much more stable cloud cover is shown in
Figure~\ref{dmResid2Y}.

\begin{figure}
\plotone{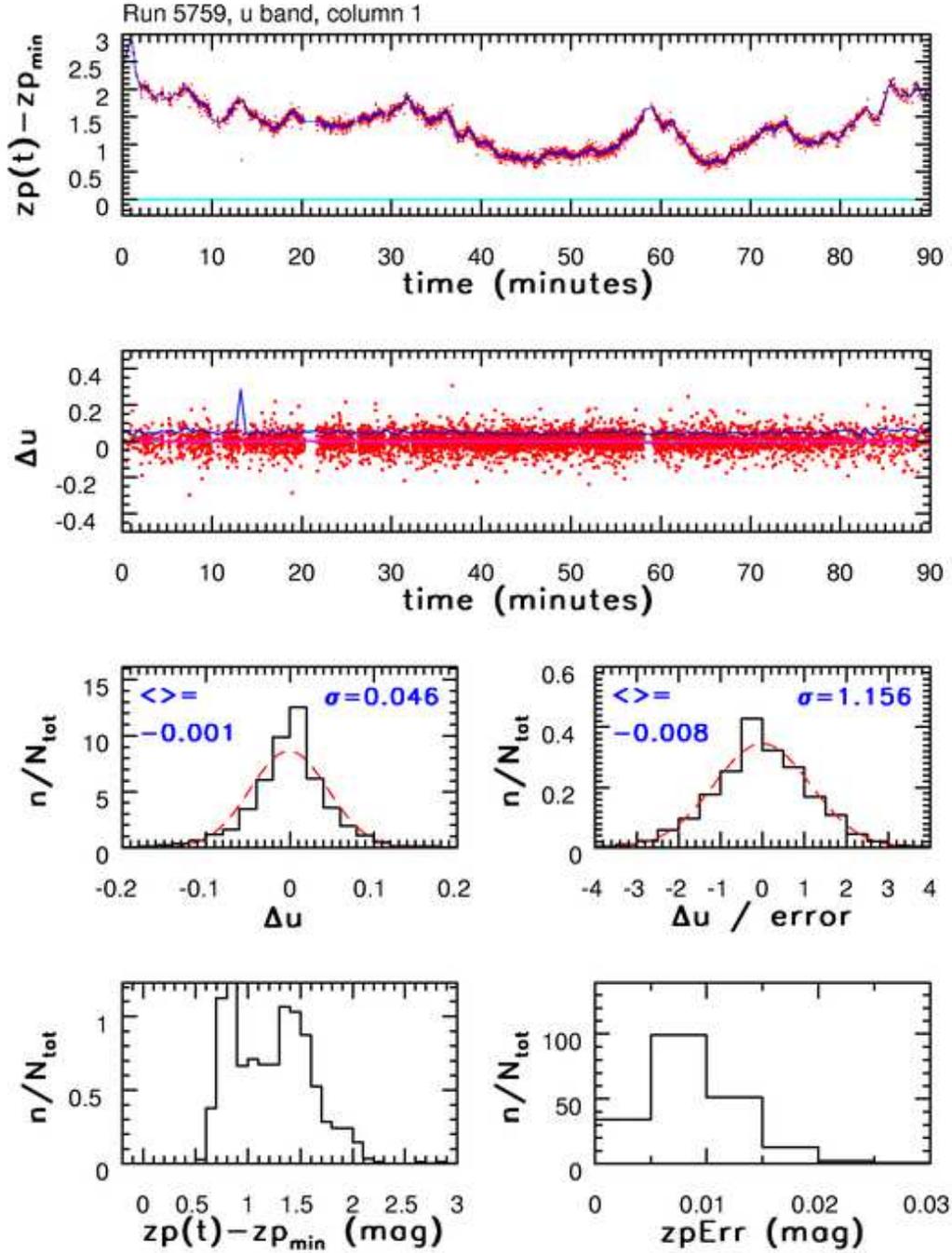}
\caption{Analogous to Fig.~\ref{dmResid2Y}, except that the $u$ band
calibration summary is shown. Despite the smaller number of calibration
stars than in redder bands, and over a magnitude of cloud extinction,
it is still possible to photometrically calibrate these data with 
a median error of only $\sim$0.01 mag. 
\label{dmResid4}}
\end{figure}

The calibration performance in other bands is similar. For example,
although the number of calibration stars is smaller in the $u$ band 
than in the $r$ band, the median zeropoint error for the same stretch
of data as shown in Figure~\ref{dmResid2Y} is still only 0.01 mag,
as illustrated in Figure~\ref{dmResid4}.

\subsubsection{ The Summary of Calibration Accuracy }
\label{summCalib}
\begin{figure}
\plotone{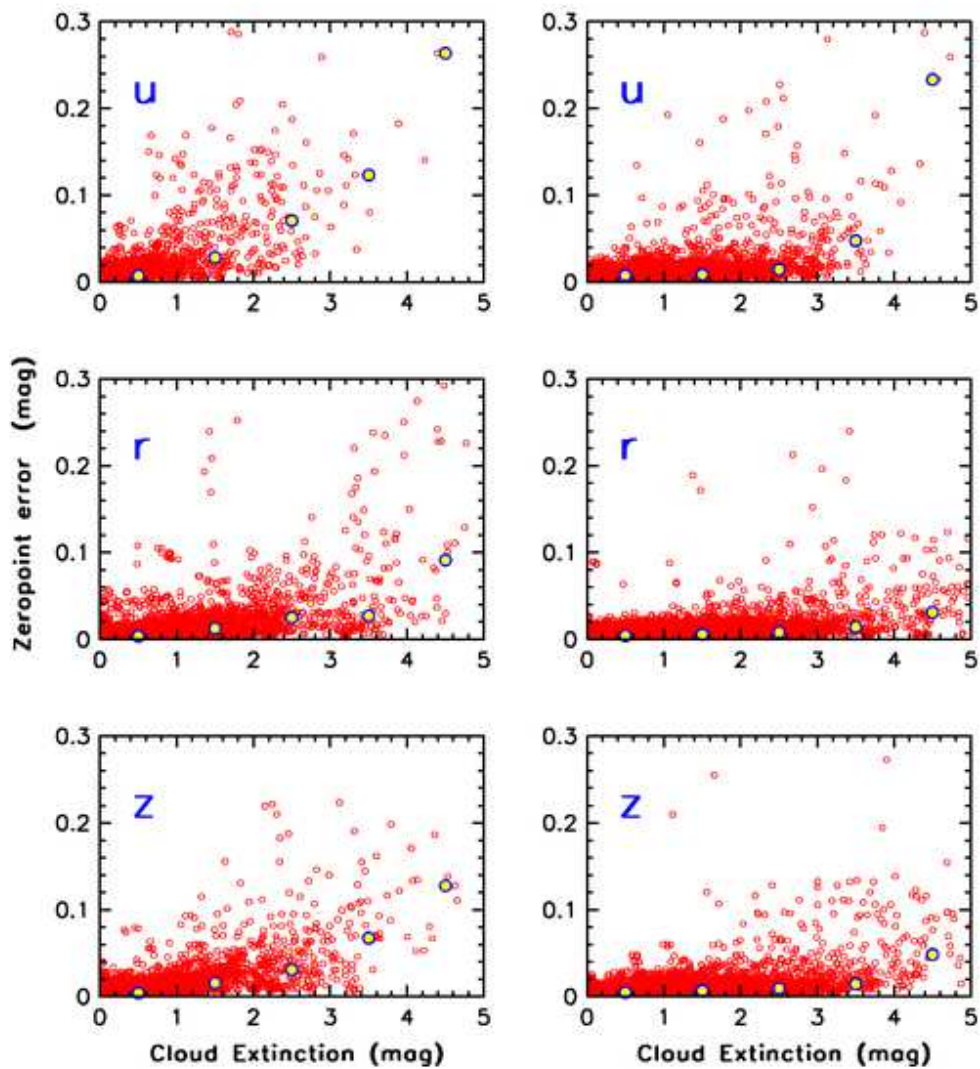}
\vskip -1.5in
\caption{A summary of calibration accuracy as a function of the
cloud extinction and band ($urz$ as marked in the panels). Each small
symbol represents one calibration patch (a $\sim$9 arcmin$^2$ large rectangle
with 1:20 aspect ratio). Zeropoint error is determined from the 
root-mean-scatter of photometric residuals. The large symbols show
the median zeropoint error in 1 mag wide bins of cloud extinction.
The left column shows data 
for one of the photometrically worst SDSS-II SNe runs (5646), and
the right column is for a run with optically thick, but exceptionally 
smooth, clouds (5759). Note that the data can be calibrated with 
zeropoint errors typically smaller than a few percent even through 
clouds several mag thick. 
\label{errVSext}}
\end{figure}

A summary of the final zeropoint errors as a function of cloud 
extinction and band for one of the worst runs is shown in the left column
in Figure~\ref{errVSext}. As the figure shows, the data can be calibrated 
with small zeropoint errors even for such a bad case. 
Typically, the zeropoint errors, for the same cloud extinction, are about 
twice as small as in this run. A calibration summary for a run with 
optically thick, but exceptionally smooth, clouds is shown in the right 
column in Figure~\ref{errVSext}. Overall, {\it for cloud extinction of 
$X$ mag, the zeropoint uncertainty is typically smaller than 
$(0.02-0.05)X$ for 95\% of calibration patches, with a median of 
(0.01-0.02)$X$.}

\subsubsection{    The Cloud Color and Structure Function  }
\label{cloudcolor}
\begin{figure}
\plotone{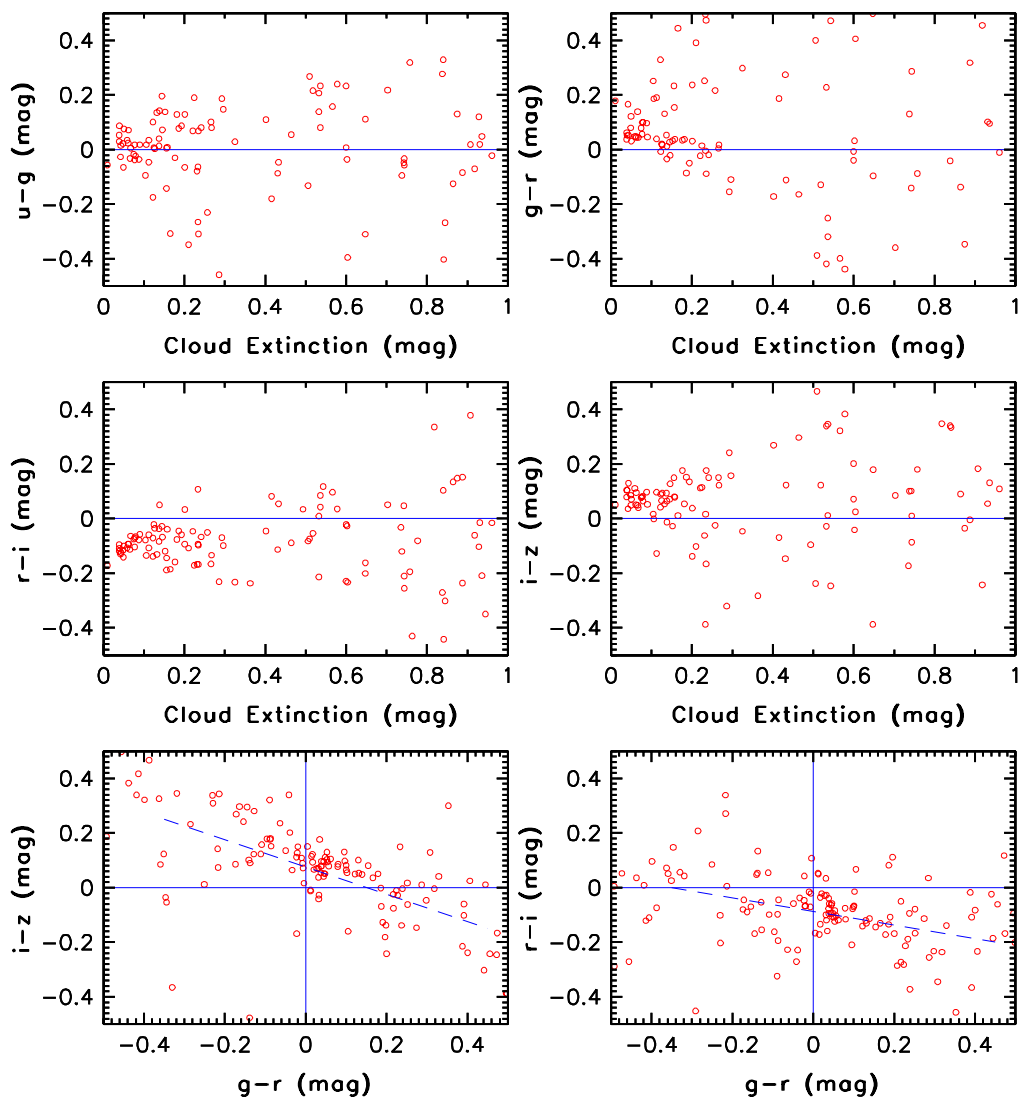}
\vskip -1in
\caption{\small The color of cloud extinction in SDSS bands. Each symbol
represents one field and shows the difference in cloud extinction
between the two bands as a function of the $r$ band extinction. 
The measurements in different bands are obtained over $\sim$5
minutes of time and thus even gray clouds with spatially varying
extinction could produce the observed non-gray (non-zero) values.
The dashed lines in the bottom two panels indicate the expected 
correlation if the color variations are due to temporal changes 
in the gray cloud thickness (rather than due to intrinsic color 
changes).}
\label{plotCloudColor}
\end{figure}

We detect no dependence of the calibration residuals on the stellar color 
or cloud thickness at a few millimag level. This is consistent with the 
lack of selective extinction by clouds. The lack of a color correlation in 
the $u$ and $z$ bands implies that the well-known cloud greyness extends beyond
visual wavelengths. Another method to quantify selective extinction by 
clouds is to directly compare zeropoints from different bands.
As shown in Figure~\ref{plotCloudColor}, the cloud extinction is similar 
in all bands for most fields. A few cases where there are deviations 
of a few tenths of a magnitude can be easily understood as due to 
temporal changes in the cloud opacity (recall that the data from 
different bands are obtained over $\sim$5 minutes of time). 

\begin{figure}
\plotone{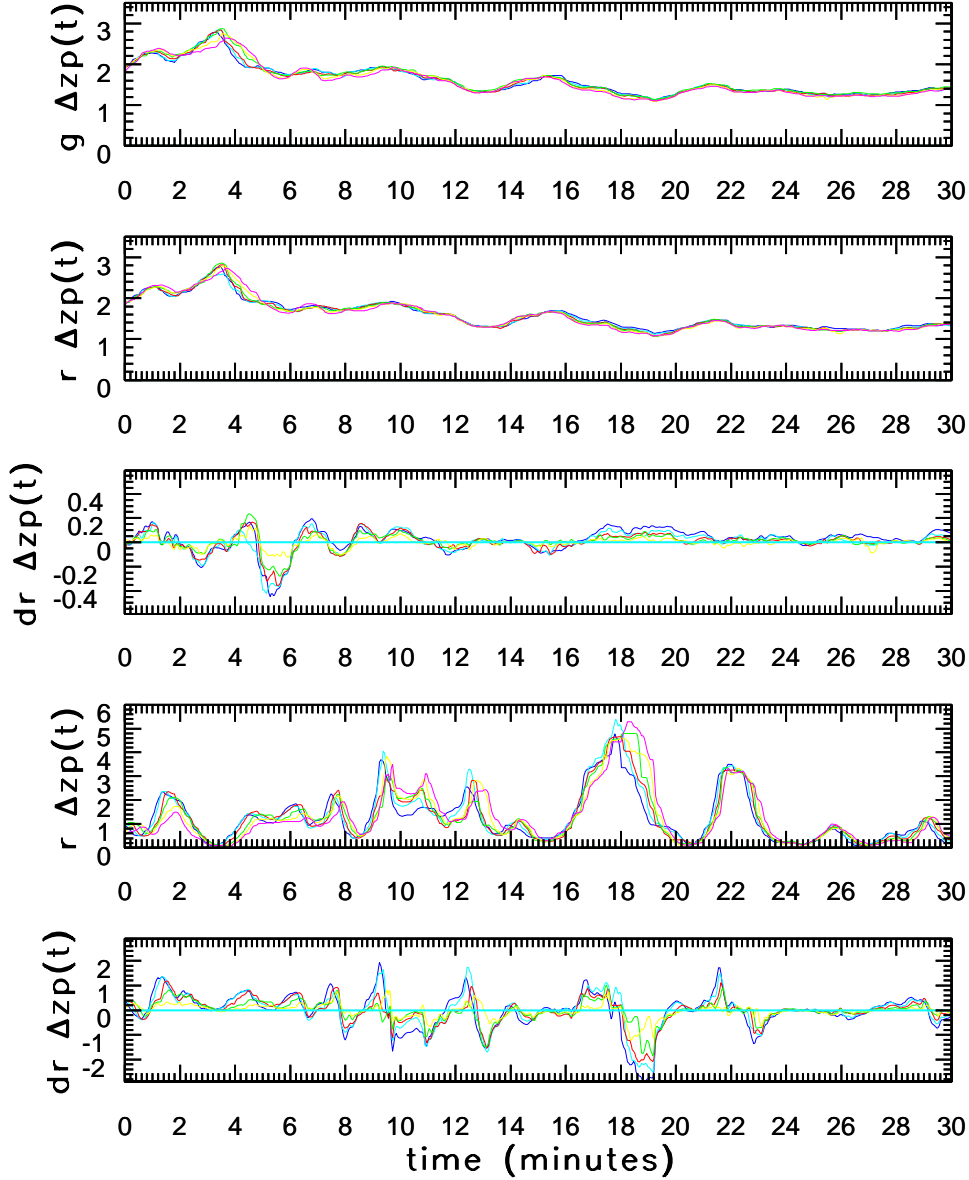}
\vspace{-1in}
\caption{\small A comparison of cloud extinction independently measured for 
six camera columns. The top two panels show the 30 minutes 
of measurements in the $g$ and $r$ band for the same run (5759)
shown in Figure~\ref{dmResid2Y}.
Individual camera columns are color-coded according to the legend
shown in the top panel in Figure~\ref{compareBandpasses}. The third
panel shows the difference of the $r$ band zeropoints measured in one 
of the edge columns (6) and zeropoints from other five columns. 
The bottom two panels are 
analogous to the second and third panel, except that the data are
from a run with exceptionally patchy clouds (5646, the first 30 minutes
of the data from Figure~\ref{dmResid1} are shown). Note the 
varying scale for the $y$ axis. 
\label{ZPBands}}
\end{figure}

\begin{figure}
\epsscale{1.2}
\plotone{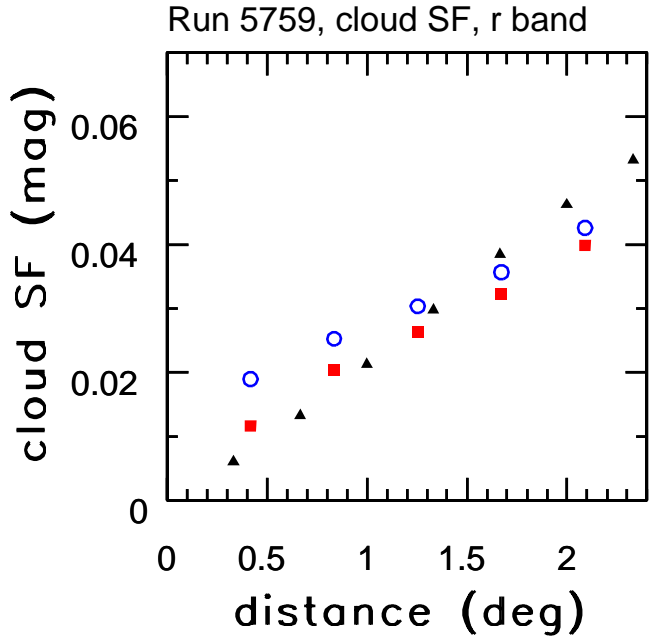}
\vspace{-2.5in}
\vskip -3in
\caption{\small The cloud structure function in the $r$ band for run 5759 
and for the same stretch of data as shown in Figure~\ref{dmResid2Y}
(the median cloud extinction is 1.3 mag). The circles show the rms 
width of the distribution of zeropoint differences between camera 
column 1 and (five) other columns. This width is corrected for a 
0.015 mag contribution from the measurement errors and shown by 
squares. The triangles show the width of the distribution 
of zeropoint differences in the in-scan direction, with
the distance scale multiplied by 30. This multiplication factor
measures the cloud speed relative to the boresight in the in-scan
direction (see \S \ref{cloudCalibration}). 
\label{CloudSF}}
\end{figure}

The calibration accuracy is determined by the size of the calibration patches. 
For example, a smaller patch would suffer less from the spatial variation 
of cloud extinction, but it wouldn't have enough stars to beat down the 
noise of their individual photometric measurements ($\sim$0.02 mag for 
sufficiently bright stars). The detailed scaling of this accuracy with 
patch size depends on the cloud spatial structure function. The 
geometry of the SDSS camera allows us to study the cloud SF on scales 
exceeding 2$^\circ$. Figure~\ref{ZPBands} compares zeropoints in different 
columns for two runs with significantly different cloud behavior. 
While zeropoints from different columns generally track each other, there 
can be differences exceeding a magnitude (they generally scale with the
cloud optical thickness). These differences increase with 
the distance between the camera columns. Figure~\ref{CloudSF} shows 
a typical behavior: for small spatial scales ($<$2$^\circ$) the SF is 
roughly a linear function of distance, and it scales roughly linearly
with the cloud extinction. {\it At a 1$^\circ$ scale, the SF is typically of 
order 2-10\% of the cloud extinction}. For example, even for clouds 3 mag thick,
the SF at 2 arcmin scales is typically  $<$0.01 mag.

\subsubsection{       Implications for Surveys such as LSST           }

The Large Synoptic Survey Telescope (LSST) is a proposed imaging survey that 
will attempt to maximize its observing time by accepting non-photometric
conditions. At the same time, it has
adopted exquisite requirements for its photometric accuracy, including
1\% accuracy of its internal photometric zeropoint errors across the
sky. Our analysis allows us to answer the following question: ``What
is the largest cloud extinction that still allows photometric
calibration accurate to 1\%?''.

A similar approach to the calibration of LSST data as presented here
(assuming that a standard star catalog is available, e.g. from prior 
demonstrably photometric nights) would benefit from several effects:
\begin{enumerate} 
\item
   The LSST will not use a drift-scanning technique and thus
   the calibration patches can be squares; for the same area,
   this results in a $\sim$5 times smaller angular scale ($\sim3$ arcmin), 
   compared to the 1:20 rectangles we have used to calibrate SDSS
   drift-scanning data. On these
   angular scales, the cloud structure function is roughly linear and 
   thus the zeropoint error would be $\sim$5 times smaller, or of
   the order 1\% or less through clouds as thick as 1 mag (conservatively
   assuming that SDSS errors would be 0.05$X$, see \S~\ref{summCalib}). We note
   that the shorter exposure time for LSST (30 seconds, or about a 
   factor of two shorter than for SDSS) is not relevant because
   clouds would typically move by more than a degree during the
   exposure. This is more than an order of magnitude larger distance 
   than the size of the calibration patch and thus the structure
   function analysis remains valid.
\item
   LSST data will be deeper than SDSS by about 2-3 mag. With a conservative
   assumption that $\log{N} \propto 0.3m$ for faint stars (0.6 for Euclidean counts),
   the surface density of calibration stars will be about ten times larger
   for LSST than for SDSS. This larger density enables
   ten times smaller patches, or about three times smaller angular
   scale for calibration ($\sim1$ arcmin), resulting in another factor 
   of three improvement of accuracy.
\item
   Fitting a smooth function for cloud opacity over several
   calibration patches would result in further improvements.
\end{enumerate} 

The first two points predict that {\it LSST data could be calibrated with 
the required 1\% accuracy even through 3 mag thick clouds}. Given the various 
extrapolations, we conservatively suggest the range of 1-3 mag as the upper 
limit on the acceptable cloud opacity\footnote{Of course, cloud opacity
decreases the imaging depth and data with clouds thicker than $\sim$1 mag
may be undesireable for reasons other than calibration accuracy.}.

\subsection{       The Morphology of the Stellar Locus  }

\begin{figure}
\epsscale{1.0}
\plotone{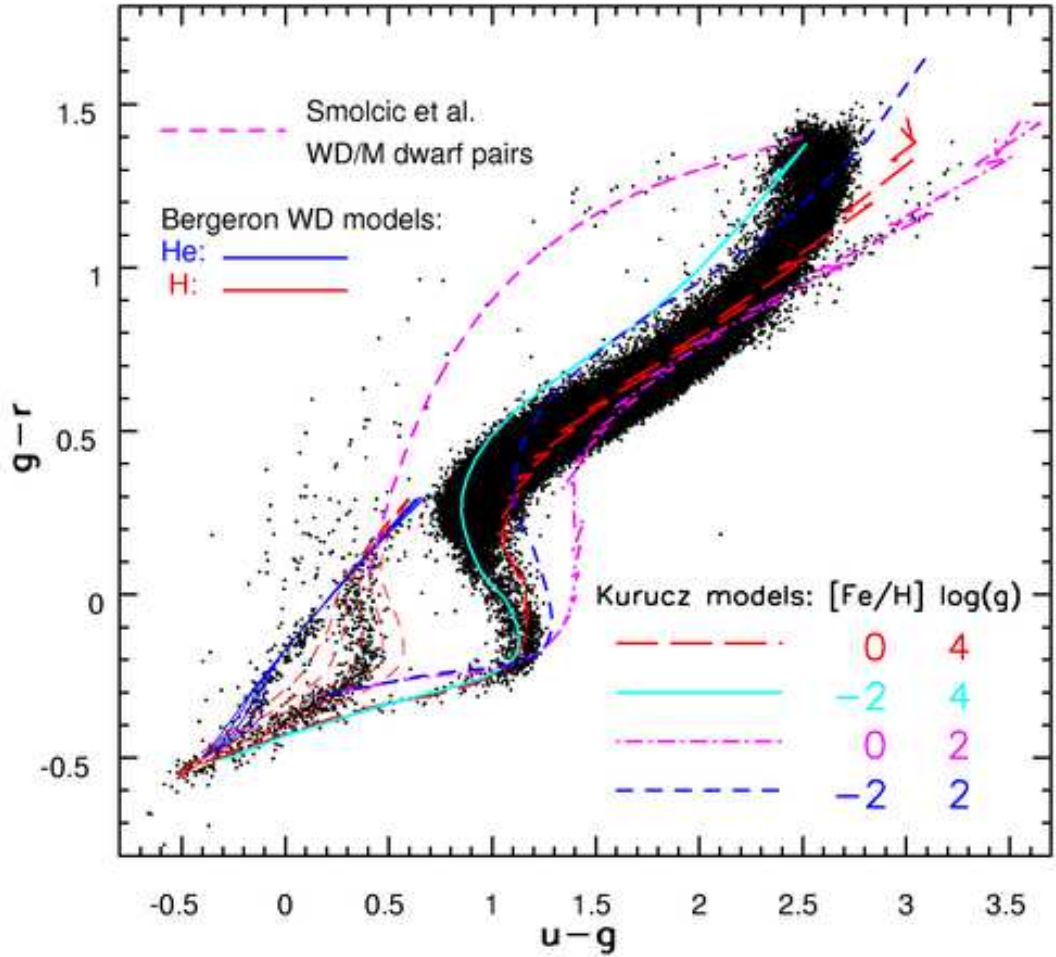}
\vspace{-2.5in}
\caption{The $g-r$ vs. $u-g$ color-color diagrams for all non-variable
point sources constructed with the improved averaged photometry (dots).
Various stellar models (Kurucz 1979; Bergeron et al. 2005; 
Smol\v{c}i\'{c} et al. 2006) are shown by lines, as indicated in the figure.
\label{superLocus}}
\end{figure}

The improved accuracy of averaged photometry provides "crisper" color-color 
diagrams and also reveals new morphological features. An example
of such a color-color diagram is shown in Figure~\ref{superLocus}. 

This is a similar plot to Figure 1 from Smol\v{c}i\'{c} et al. (2004),
except that {\it only} non-variable point sources are shown (note the 
absence of quasars) and averaged photometry is used. The white 
dwarf/M dwarf  "bridge" discussed by Smol\v{c}i\'{c} et al. is clearly 
visible, as well as the locus of probable solar metallicity giants (
this identification is based on models, e.g. Kurucz 1979) which bifurcates 
from the main locus at $u-g \sim 2.5$ and $g-r\sim 1$. Note also the 
well-defined BHB locus ($u-g \sim 1.1$ and 
$g-r$ from $-$0.3 to 0.1) and the white dwarf locus ($u-g \sim 0.35$ and 
$g-r$ from $-$0.3 to $\sim$0.0). A new locus-like feature, that is {\it not} 
visible in Figure 1 from Smol\v{c}i\'{c} et al., is discernible at 
$u-g \sim 0$ and $g-r \sim -0.2$. The great value of the accurate
$u$ band photometry  is clearly evident at e.g. $g-r =-0.2$: the $u-g$ 
color distribution is tri-modal!  The bluest branch is consistent with 
He white dwarfs, and the middle branch with hydrogen white dwarfs, as 
supported by Bergeron et al. (1995) white dwarf models and detailed
analysis of SDSS spectra (Eisenstein et al. 2006). The reddest 
branch is made of blue horizontal branch stars (see Sirko et al. 2004
and references therein). 

\begin{figure}
\plotone{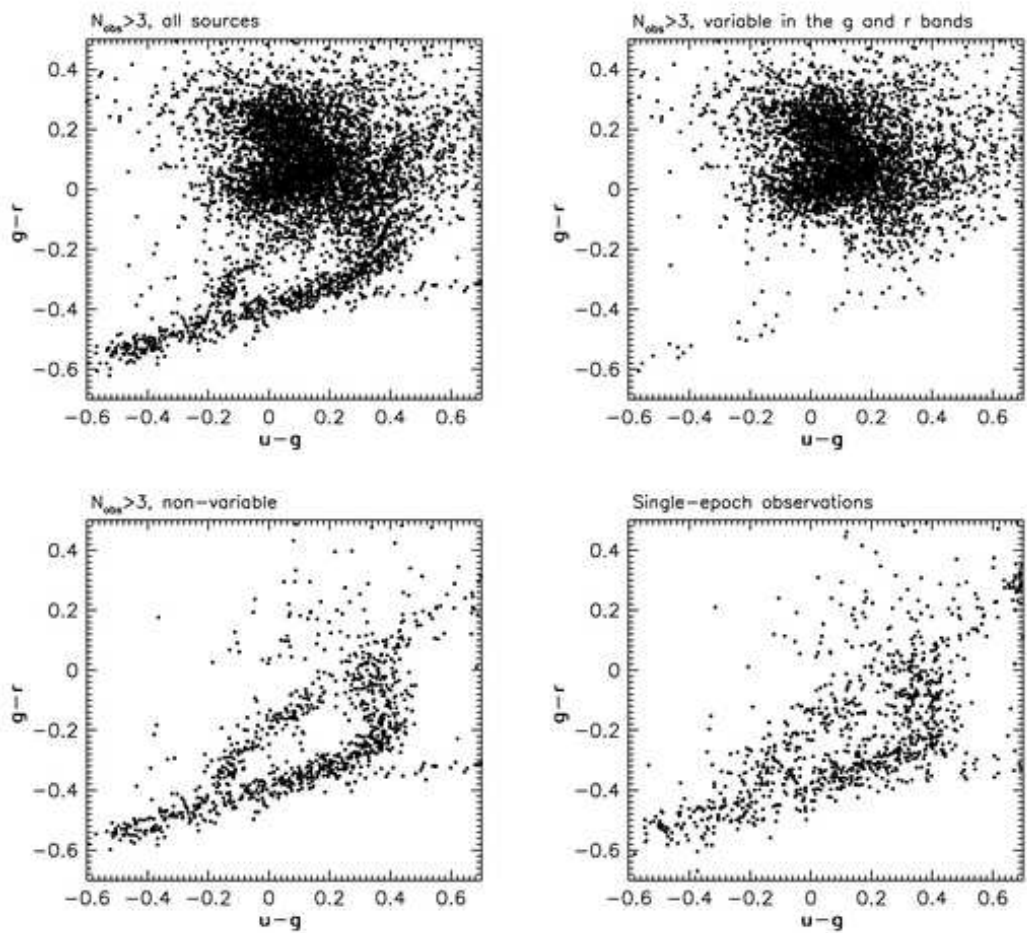}
\vspace{-1.5in}
\caption{An illustration of the advantages of repeated 
photometric measurements. The top left panel shows the blue corner 
of the $g-r$ vs. $u-g$ diagram from Figure~\ref{superLocus} for 
{\it all} point sources with the averaged photometry. The top right 
panel shows {\it only} the variable sources (dominated by low-redshift 
quasars), and the bottom left panel shows the non-variable sources
(dominated by white dwarfs), classified using low-order lightcurve
moments. The bottom right panel shows the {\it same} non-variable 
sources, but using their DR5 single-epoch photometry. A comparison
of the bottom two panels shows the striking improvement made
possible by the use of multiple observations of the same field.
\label{lsstPreview}}
\end{figure}

The exciting fact that {\it one can distinguish He and H white dwarfs
using photometry alone} is a consequence of the improved 
photometric accuracy due to averaging many epochs.
Figure~\ref{lsstPreview} reiterates that point. Note the striking 
difference between the two bottom panels: while one could be convinced 
that the He white dwarf sequence is a real feature in the bottom 
right panel, its existence is clearly evident when using the improved 
photometry, as shown in the bottom left panel. In summary, {\it the 
multi-epoch observations provide both the identification of variable 
sources and much more accurate colors for non-variable sources.}
This bodes well for science deliverables from upcoming large-scale 
imaging surveys. For example, LSST will obtain over its 10-year
long mission similar repeat imaging as discussed here, but about 
2.5 mag deeper, with about 100 or more observations per band and 
object, and over about two orders of magnitude larger area. Although
these new surveys will not have a spectroscopic component like SDSS
did, the multi-epoch nature of their imaging will provide alternative
information-rich datasets.

\section{               Discussion and Conclusions                             }

Using repeated SDSS measurements, we have constructed a catalog of 
over a million candidate standard stars. The catalog is publicly
available from the SDSS website\footnote{See
http://www.sdss.org/dr5/products/value\_added/index.html}. 
Several independent tests suggest that both internal zeropoint errors 
and random photometric errors for stars brighter than (19.5, 20.5, 20.5, 20, 
18.5) in $ugriz$, respectively, are at or below 0.01 mag (about 2-3 times better 
than for an individual SDSS scan). This is by far the largest existing catalog 
with multi-band optical photometry accurate to $\sim$1\%, and breaks the 
accuracy barrier discussed by e.g. Stubbs \& Tonry (2006, and references therein). 
These observations were not obtained for the specific purpose of calibration,
but were part of the regular SDSS observational program. When compared to, for example, 
the heroic calibration efforts by Landolt, Stetson, and others, it seems justified 
to call the method presented here ''industrial'' photometry. However,
the catalog presented here is not without its problems. 

The selection of candidate stars was simply based on the absence of variability.
It is fairly certain that not all variable sources are recognized 
because of the limited number of repeated observations ($\sim$10). For example, 
an eclipsing binary with much shorter eclipse duration than the orbital
period could easily escape detection. Furthermore, some of these sources 
may not even be stars. A cross-correlation with the SDSS spectroscopic 
database yields 99,000 matches in the overlapping region, including 44,000 
spectra classified as stars. About 70 candidate standard stars are actually 
spectroscopically confirmed quasars! Apparently, a small fraction 
of quasars (a few percent, for a detailed analysis see Sesar et al. 2007, in
prep.) cannot be detected by variability (at least not using
the number of epochs, their time distribution, and photometric accuracy employed
in this work). Indeed, we have also found three spectroscopically confirmed SDSS
quasars among Stetson's standards which were observed 20-30 times and which
showed no variation. Similarly, about 300 candidate standard stars have SDSS 
spectra classified as galaxies. Nevertheless, the inspection of color-color 
diagrams strongly suggest that the overwhelming majority of the standard stars 
are found on the stellar locus. 

Remaining systematic errors are another important concern. Effectively, we 
have assumed that PT problems average out in many patches when deriving 
flatfield corrections using stellar colors. This may not be true at a level 
not much smaller than 1\%, and thus the remaining gray problems at such a
level may be present in the catalog. Despite these residual problems, we
believe that internal consistency of the catalog (i.e. when ignoring
$\Delta_m$ from eq.~\ref{photoSysErr}) is such that the rms width for the 
function $\delta_m({\rm RA,Dec})$ from eq.~\ref{photoSysErr} evaluated for all 
stars in the catalog is at most 0.01 mag in the $griz$ bands and perhaps just 
slightly larger in the $u$ band (very unlikely exceeding 0.02 mag).
In addition to gray problems and overall flatfield errors, the dependence of 
flatfields on source color is probably the largest remaining systematic error.  

We illustrate several use cases for this catalog, including the calibration of 
highly non-photometric data and robust selection of stars with peculiar colors. 
We find that LSST and similar surveys will be able to observe in partially cloudy
(non-photometric) nights because even cloudy data can be accurately 
calibrated with a sufficiently dense network of calibration stars. Such a dense 
network will be self-calibrated by LSST very soon after first light, 
using an approach developed for SDSS data by Padmanabhan et al. (2007).  
Given such a network, SDSS experience suggests that LSST can maintain its 
required photometric calibration accuracy of 1\% even when observing 
through 1-3 mag thick clouds.

Perhaps the most exciting conclusion of this work is that it may 
become obsolete in only a few years due to the advent of next-generation
surveys such as Pan-STARRS and LSST.

\vskip 0.4in \leftline{Acknowledgments}

We are thankful to Peter Stetson for his most useful comments and for
providing  his latest standard star photometry to us. \v{Z.} Ivezi\'{c}
and B. Sesar acknowledge support by NSF grant AST-0551161 to LSST for design 
and development activity, and H. Morrison acknowledges support by NSF grant 
AST-0607518.
    
Funding for the SDSS and SDSS-II has been provided by the Alfred P. Sloan Foundation, the Participating Institutions, the National Science Foundation, the U.S. Department of Energy, the National Aeronautics and Space Administration, the Japanese Monbukagakusho, the Max Planck Society, and the Higher Education Funding Council for England. The SDSS Web Site is http://www.sdss.org/.

The SDSS is managed by the Astrophysical Research Consortium for the Participating Institutions. The Participating Institutions are the American Museum of Natural History, Astrophysical Institute Potsdam, University of Basel, University of Cambridge, Case Western Reserve University, University of Chicago, Drexel University, Fermilab, the Institute for Advanced Study, the Japan Participation Group, Johns Hopkins University, the Joint Institute for Nuclear Astrophysics, the Kavli Institute for Particle Astrophysics and Cosmology, the Korean Scientist Group, the Chinese Academy of Sciences (LAMOST), Los Alamos National Laboratory, the Max-Planck-Institute for Astronomy (MPIA), the Max-Planck-Institute for Astrophysics (MPA), New Mexico State University, Ohio State University, University of Pittsburgh, University of Portsmouth, Princeton University, the United States Naval Observatory, and the University of Washington.

\newpage

\begin{deluxetable}{rrrrrrrrrrr}
\tablenum{1} \tablecolumns{11} \tablewidth{400pt}
\tablecaption{The Color-term Corrections$^a$}
\tablehead{column & $\left\langle du \right\rangle$ & $\sigma_{du}$ &
$\left\langle dg \right\rangle$ & $\sigma_{dg}$ &
$\left\langle dr \right\rangle$ & $\sigma_{dr}$ &
$\left\langle di \right\rangle$ & $\sigma_{di}$ &
$\left\langle dz \right\rangle$ & $\sigma_{dz}$}
\startdata
      1  &  1.0 &  0.3 &  1.8 &  0.4 & -1.2 &  0.3 &  0.0 &  0.2 &  0.6 &  0.5 \\
      2  & 11.0 &  2.3 &  7.6 &  1.6 & -1.1 &  0.2 & -0.0 &  0.2 & -0.8 &  1.3 \\
      3  & -5.4 &  0.8 & -2.0 &  0.0 &  2.9 &  1.0 & -0.2 &  0.3 & -0.2 &  0.5 \\
      4  & -9.5 &  2.1 & -1.0 &  0.2 & -0.9 &  0.1 &  0.0 &  0.0 & -0.4 &  0.3 \\
      5  &  3.9 &  1.0 & -4.4 &  1.3 & -0.0 &  0.2 & -0.2 &  0.1 &  0.4 &  0.7 \\
      6  & -0.5 &  0.8 & -2.6 &  0.4 & -0.6 &  0.1 & -0.2 &  0.3 &  0.9 &  1.2 \\
\enddata
\tablenotetext{a}{The median and rms scatter for photometric
corrections that place the measurements in six camera columns on 
the survey system (in millimag). } 
\end{deluxetable}

\begin{deluxetable}{ccccccc}
\tablenum{2} \tablecolumns{5} \tablewidth{400pt}
\tablecaption{The $s, w, x, y$ Principal Color Definitions}
\tablehead{$PC$  &    A   &    B   &    C   &    D   &    E   &   F }
\startdata
   s   & -0.249 &  0.794 & -0.555 &  0.0   &  0.0   &  0.234 \\ 
   w   &  0.0   & -0.227 &  0.792 & -0.567 &  0.0   &  0.050 \\
   x   &  0.0   &  0.707 & -0.707 &  0.0   &  0.0   & -0.988 \\
   y   &  0.0   &  0.0   & -0.270 &  0.800 & -0.534 &  0.054 \\
\enddata
\end{deluxetable}

\begin{deluxetable}{ccrcc}
\tablenum{3} \tablecolumns{5} \tablewidth{400pt}
\tablecaption{The Effect of Repeated Measurements on the Width of the Stellar Locus}
\tablehead{PC & rms 1 obs$^a$ & rms N obs$^b$ & median$^c$  & width for PC/PCerror$^d$}
\startdata
   $s$  &    31    &     19  &   3.0   &     3.02 \\
   $w$  &    25    &     10  &   1.1   &     1.82 \\
   $x$  &    42    &     34  &   1.2   &     5.34 \\
   $y$  &    23    &      9  &   0.8   &     1.64 \\  
\enddata
\tablenotetext{a}{The locus width determined using single-epoch SDSS observations (in millimag).} 
\tablenotetext{b}{The locus width determined using multiple SDSS observations (in millimag).} 
\tablenotetext{c}{The median principal color determined using multiple SDSS observations (in millimag).} 
\tablenotetext{d}{The locus width normalized by expected measurement errors.} 
\end{deluxetable}

\begin{deluxetable}{cccc}
\tablenum{4} \tablecolumns{5} \tablewidth{400pt}
\tablecaption{The Flatfield Corrections$^a$}
\tablehead{band & width$^b$ & min$^c$ & max$^d$}
\startdata
     $ur$   &    22 & -53 & 53 \\
     $gr$   &    12 & -27 & 19 \\
     $r$    &     7 & -17 & 17 \\
     $ri$   &     4 & -10 & 13 \\
     $rz$   &     7 & -14 & 19 \\
\enddata
\tablenotetext{a}{The $r$ band correction is determined using observations by the Photometric
                  Telescope, and the $ugiz$ corrections are determined using the stellar locus
                  method (see \S \ref{PCmethod} and \ref{PTmethod}).} 
\tablenotetext{b}{The root-mean-square scatter for applied flatfield corrections (in millimag).} 
\tablenotetext{c}{The minimum value of the applied correction (in millimag).} 
\tablenotetext{d}{The maximum value of the applied correction (in millimag).} 
\end{deluxetable}
     
\begin{deluxetable}{rrrrrrr}
\tablenum{5} \tablecolumns{5} \tablewidth{400pt}
\tablecaption{The statistics of the median PT-2.5m residuals}
\tablehead{band & $<>^a$ & width$^b$ & min$^c$ & max$^d$ & N$^e$ & $<>^f$ }
\startdata
   $u$  &  -2  &  9  &  -18  &  23   &   175     &         5 \\
   $g$  &   6  &  7  &   -4  &  17   &   647     &         5 \\ 
   $r$  &   3  &  7  &   -7  &  10   &   627     &         3 \\
   $i$  &   4  &  7  &  -10  &  17   &   621     &         2 \\
   $z$  &   1  &  8  &  -16  &  15   &   286     &        -2 \\
\enddata
\tablenotetext{a}{The median value for the bin medians (in millimag). There are 24 bins, distributed
   inhomogeneously in the right ascension direction.} 
\tablenotetext{b}{The root-mean-square scatter for the bin medians  (in millimag).} 
\tablenotetext{c}{The minimum value for the median residuals (in millimag).} 
\tablenotetext{d}{The maximum value for the median residuals (in millimag).} 
\tablenotetext{e}{The median number of stars per bin.} 
\tablenotetext{f}{The median value of the residuals for stars with colors within 
                  0.02 mag from the crossing colors (in millimag).} 
\end{deluxetable}

\begin{deluxetable}{crrrr}
\tablenum{6} \tablecolumns{5} \tablewidth{400pt}
\tablecaption{Residuals from the mean color-redshift relation$^a$ for red galaxies.}
\tablehead{color & rms(Dec)$^b$ & $\chi$(Dec) & rms(RA)$^c$ & $\chi$(RA) }
\startdata
     $u-g$   &   21 &  1.5  &  18 &  1.3  \\
     $g-r$   &    4 &  1.3  &  12 &  2.6  \\
     $r-i$   &    3 &  1.3  &   6 &  3.5  \\
     $i-z$   &    9 &  2.9  &   6 &  2.9  \\
\enddata
\tablenotetext{a}{The table lists the rms widths of color-residual distributions 
(in millimag), and the widths of distributions of residuals normalized by 
statistical noise ($\chi$), using mean color-redshift relations (see text).}
\tablenotetext{b}{The rms for the declination direction, using 0.025$^\circ$ wide bins (0.1$^\circ$
  for the $u-g$ color).} 
\tablenotetext{c}{The rms for the right ascension direction, using 2$^\circ$ wide bins (5$^\circ$
  for the $u-g$ color).} 
\end{deluxetable}

\begin{deluxetable}{crrrrrrrrr}
\tablenum{7} \tablecolumns{5} \tablewidth{500pt}
\tablecaption{Comparison with Stetson's standards: I. {\it griz} to {\it BVRI} transformations$^a$}
\tablehead{color & $<>_{med}^b$ & $\sigma_{med}^c$ & $\chi_{med}^d$ & $<>_{all}^e$ & $\sigma_{all}^f$ & A$^g$ & B$^g$ & C$^g$ & D$^g$}
\startdata
$B-g$ & -1.6 & 8.7 & 1.4 &  1.0 &  32 &  0.2628 & -0.7952 &  1.0544 &  0.0268  \\
$V-g$ &  0.8 & 3.9 & 1.0 &  0.9 &  18 &  0.0688 & -0.2056 & -0.3838 & -0.0534  \\
$R-r$ & -0.1 & 5.8 & 0.9 &  1.2 &  15 & -0.0107 &  0.0050 & -0.2689 & -0.1540  \\ 
$I-i$ &  0.9 & 6.1 & 1.0 &  1.2 &  19 & -0.0307 &  0.1163 & -0.3341 & -0.3584  \\
\enddata
\tablenotetext{a}{These transformations are valid {\it only} for main sequence stars with 
colors in the range $g-r>0.2$ and $r-i<1.5$ (roughly, 0.3$< B-V <$1.6). Extrapolation outside 
this range may result in large (0.1 mag or larger for hot white dwarfs) errors!}
\tablenotetext{b}{The median value of median transformation residuals (differences between the measured 
values of colors listed in the first column and those synthesized using eq.~\ref{BVRItrans}) in 0.1 mag 
wide $g-r$ bins for stars with 0.25$< g-r <$1.45 (in millimag). These medians of medians measure the
typical level of systematics in the $gri$-to-$BVRI$ photometric transformations introduced by the adopted
analytic form (see eq.~\ref{BVRItrans}).} 
\tablenotetext{c}{The root-mean-square scatter for median residuals described above (in millimag).} 
\tablenotetext{d}{The root-mean-square scatter for residuals normalized by statistical noise.
The listed values are $\sim$1, which indicates that the scatter around adopted photometric transformations
listed under b) is consistent with expected noise.} 
\tablenotetext{e}{The median value of residuals evaluated for all stars (in millimag).} 
\tablenotetext{f}{The root-mean-square scatter for residuals evaluated for all stars (in millimag).} 
\tablenotetext{g}{Coefficients A--D needed to transform SDSS photometry to the BVRI system (see eq.~\ref{BVRItrans}).} 
\end{deluxetable}

\begin{deluxetable}{crrrrrrrrrrrr}
\tablenum{8} \tablecolumns{5} \tablewidth{500pt}
\tablecaption{Comparison with Stetson's standards: II. Photometric zeropoint variations}
\tablehead{color & $<>_{R1}^a$ & $\sigma_{R1}^b$ & N$_{R1}^c$ & $<>_{R2}^a$ & $\sigma_{R2}^b$ & N$_{R2}^c$ & $<>_{R3}^a$ & $\sigma_{R3}^b$ & N$_{R3}^c$ & $<>_{R4}^a$ & $\sigma_{R4}^b$ & N$_{R4}^c$}
\startdata
 $B-g$ &     $-$29 &  21 &   92 &      6 &  27 &  165 &      8 &  42 &  155 &   $-$4 &  27 &  281 \\
 $V-g$ &       0   &  17 &   99 &      0 &  15 &  217 &      6 &  25 &  161 &     17 &  19 &  282 \\
 $R-r$ &     $-$6  &  16 &   58 &      4 &  16 &  135 &   $-$8 &  12 &   11 &     39 &  27 &   60 \\
 $I-i$ &     $-$11 &  16 &   94 &      6 &  18 &  205 &      2 &  16 &  124 &     19 &  15 &   47 \\
\enddata
\tablenotetext{a}{The median value of residuals (in millimagnitudes) for transformations listed in the first 
column, evaluated  separately for regions 1-4, defined as: 
{\bf R1:} RA$\sim$325, Dec$<$0; {\bf R2:} RA$\sim$15; {\bf R3:} RA$\sim$55;
 {\bf R4:} RA$\sim$325, Dec$>$0.} 
\tablenotetext{b}{The root-mean-square scatter for the transformation residuals (in millimagnitudes).} 
\tablenotetext{c}{The number of stars in each region with good photometry in the required bands.} 
\end{deluxetable}

\newpage

\clearpage

\clearpage

\newpage

\clearpage

\newpage

\newpage

\newpage

\end{document}